\documentclass[preprint,preprintnumbers,amsmath,amssymb]{revtex4}
\usepackage[colorlinks=true, pdfstartview=FitV, linkcolor=blue, citecolor=red, urlcolor=magenta]{hyperref}
\usepackage[utf8]{inputenc}
\usepackage{amssymb}
\usepackage{amssymb}
\usepackage{amsmath}
\usepackage{latexsym}
\usepackage{epsfig}
\usepackage{color}
\allowdisplaybreaks
\begin{document}
\title{\textbf{Thermal fluctuations, quasi-normal modes and phase transition of the charged AdS black hole with perfect fluid dark matter}}
\author{ G
Abbas $^1$ \footnote{ghulamabbas@iub.edu.pk }, R. H. Ali $^1$
\footnote{hasnainali408@gmail.com}}
\address{ $^1$ Department of Mathematics, The Islamia University of Bahawalpur,\\ Bahawalpur Pakistan}

\date{}

\begin{abstract}

In this paper, we study thermodynamics, thermal fluctuations, phase transitions and the charged anti-de Sitter black hole surrounded by perfect fluid dark matter. Large black holes are shown to be stable when subject to thermal fluctuations, and we begin by exploring how these fluctuations affect the uncorrected thermodynamic quantities of entropy, Helmholtz free energy, Gibbs free energy, enthalpy specific heat, and phase transition stability. We also discuss null geodesics and the radius of the photon sphere for the charged AdS BH and use the radius of a photon sphere to calculate the Lyapunov exponent and angular velocity. Exceptionally, we test the effects of various parameters of a black hole graphically by observing the existence of the correction parameter and the coupling parameter, which reveal the behavior of corrected thermodynamic quantities. Lastly, we see how the system is stable (under the effects of the dark matter parameter) by figuring out the specific heat and Hawking temperature, which are both related to entropy.
\end{abstract}

\maketitle
\section{Introduction}
One of the most captivating subjects in General Relativity (GR) is the study of the aftereffects of self-gravitating objects. Black hole (BH) is a thermodynamically unstable object with the complete destructive structure of massive stars, having the strongest gravitational force as nothing can get away from it. Black hole as a thermodynamical system provides us with a platform in the field of gravitational physics. The formulation of four laws of BH mechanics was investigated by Bardeen, Carter and Hawking \cite{1}.
Bekenstein \cite{2} and Hawking \cite{3,4} are the pioneers who made impressive developments in the field of BH thermodynamics. Black holes exhale subatomic particles at the microscopic level, and these subatomic particles are known as Hawking radiations, which tells us about BH geometry. Modifying the Bekenstein entropy relation is necessary for introducing the holographic principle \cite{5} and thermal fluctuation.
Due to quantum fluctuations in statistical mechanics, which are in fact thermal fluctuations around equilibrium, the maximum entropy of BHs would be corrected when their size decreases as a result of Hawking radiation. In thermodynamics, the entropy of a system is seen to be a logarithmic correction at the first order by minor statistical perturbations near equilibrium. Quantum fluctuations in the BH geometry cause the thermal fluctuations in its thermodynamics. The corrected thermodynamics of BHs bring local and global stability or instability, criticality, holographic duality, and many other key properties to BHs. Thermodynamical corrections by thermal fluctuations are important in BH physics. This establishes a link between thermodynamics and space-time geometry. Due to the change in thermal fluctuation \cite{6} the area-entropy relation needs to be changed by some logarithmic factor. The procedure for correcting entropy at the equilibrium state was investigated by Das et al. \cite{7}.
The logarithmic correction term exists only when Schwarzschild BH can be taken in the cavity of a small radius \cite{8} and cannot be applied when it is thermally unstable and evaporated by Hawking radiations. There are two approaches to proving the existence of the logarithmic correction term: the density of the state \cite{9} and the Cardy formula \cite{10}, developed in the context of conformal field theory (CFT). The significance of these corrected logarithmic terms has been observed for the various BHs geometries in the literature \cite{11}. Investigating the role of correction terms of higher order on the five-dimensional extreme BH with the help of supersymmetry was examined by Alishahiha \cite{12}.

Sadeghi et al. \cite{13} have investigated how the Hawking-Page transition is affected by the logarithmic correction to hyper-scaling. The outcome of the quantum correction on the charged and amended Hayward BHs is discussed by Pourhassan and Fazal \cite{14}. Shahzad and Jawad \cite{15} also review the consequences of modified entropy for non-minimal regular BHs, and it is found that under these logarithmic correction terms, the stability of BHs increases for non-minimal regular cases. In the same scenario, Pourdavish et al. \cite{16} describe the thermal properties of G\"{o}del BH, which calculate the partition function along with the probability of the observed state. Pourhassan et al. \cite{17} study the first law of thermodynamics as well as the criticality and instability of BHs when there is a higher order of correction.

In the literature \cite{18} the higher-dimensional BHs were discussed under the influence of thermal fluctuation. Thermodynamical stability was examined by Holdar et al. \cite{19}, which contain Gibbs energy, Helmholtz free energy, and the enthalpy of BH in the Lovelock theory under AdS and regular BHs. Under the influence of first-order correction, Zhang \cite{20} investigate the thermodynamical potential and phase transition of AdS-RN and Kerr-Newmann BHs. Pourhassan and Upadhyay \cite{21} evaluated the effectiveness of the thermal correction for the stability and phase change of AdS-charged BH. Basically, quasi-normal modes (QNMs) are the perturbed solution that distinguishes BHs from other compact objects. Vishveswara \cite{22} was the first to initiate the analytical depiction of the QN wave function to demonstrate the analyticity of the static and rotating BHs. Jing and Pan \cite{23} introduced a way for finding the correlation between the QNMs and the RN-BH phase changes.

He et al.\cite{24} have investigated the association for QNMs and charged Kaluza-Klein BH with QN perturbation. Konoplya and Zhidenko \cite{25} have revealed the many properties of BH perturbation, such as the removal of quantities in the perturbation and the holographic superconductor. Konoplya and Stuchlik \cite{26} have adopted the unique procedure named the \textit{eikonal method}, which is satisfactory for studying the QNMs of BH as well as null geodesics. Breton et al. \cite{27} computed the massless scaler field for QNMs by adopting the Wentzel-Kramers-Brillouis approximation and explored the radial null geodesics by using the eikonal method. In the same scenario, Churilova \cite{28} has explored the analytical solution by the eikonal method for QNMs in asymptotically flat space-time. Recently, Sharif and Zunera \cite{29} have clarified the thermodynamic characteristics of charged BHs and explained that, due to the effect of logarithmic correction, the small-radius BHs collapse to singularity.
Latter on, Sharif and Amjad \cite{30} illustrated the thermal fluctuations with the uncorrected quantities of thermodynamics, that is, the entropy, the temperature, the Helmholtz free energy, the Gibbs free energy, the enthalpy, the internal energy, the heat capacity, and the thermal stability. They found that massive BHs are more stable under the influence of thermal fluctuation.

The goal of this article is to investigate how perfect fluid dark matter (PFDM) affects BH solutions and the known thermodynamic properties of the BH solution with spherical symmetry in AdS space-time. Black hole chemistry is the study of BH thermodynamics in conjunction with the presence of a cosmological constant considered as its pressure. Our purpose is to link the idea of BH chemistry \cite{31, 32} with charged BHs in PFDM. There has been a lot of progress in the observational fields of astrophysics and astronomy in recent years. The majority of the mass-energy in the cosmos is made up of $73\%$ DE \cite{33}, $23 \%$ DM \cite{34, 35} and then $4\%$ baryonic matter according to the standard model of cosmology, as confirmed by observations such as the cosmic microwave background (CMB), baryon acoustic oscillations (BAO), weak lensing, Type Ia supernovae, etc. Several new theoretical models of DE and DM have been conceived \cite{36, 37}, to account for the aforementioned observations. The literature on DE models includes the quintessence, phantom, and quintom models. In many works, the ratio of the pressure to the energy density of DE \cite{38, 39, 40} serves as the defining feature of the quintessence model. So, it makes sense to investigate how quintessence interacts with BHs. This corresponds to the BH solution proposed by Kiselev \cite{A41}, which depicts the Schwarzschild BH surrounded by the quintessence. Many subsequent works have been conducted to investigate the BH in the quintessence field \cite{A42}-\cite{A48}.

Cold dark matter (CDM), scalar field dark matter (SFDM) and warm dark matter (WDM) are suitable examples of DM models\cite{A49}-\cite{A53}. Perfect fluid dark matter (PFDM) is a new model of DM that has attracted a lot of interest over the past two decades. Since the PFDM model was able to explain why the rotation curves of spiral galaxies are asymptotically flat \cite{A54}, it has given a whole new area to study for the astrophysical community. The Event Horizon Telescope (EHT) recently showed pictures of $M87*$ and Sgr $A*$, two super-massive BHs. The EHT observations can put limits on strange and new physical processes. In particular these can suggest or rule out certain DM models. Dark matter is a physical object whose gravitational effects have been seen in many different systems but its true nature is still mystery \cite{A55}-\cite{A58}.
% It is an important part of modern cosmology. The BHs with PFDM model have been presented in .  
 
The purpose of this paper is to examine the correction term of entropy through thermal fluctuation for the AdS BH with the PFDM. Furthermore, we have discussed the QNMs, stability, and the phase transition of AdS BH with the PFDM. The paper is organized as follows: In the section \textbf{II}, we explain the solution of RN-AdS BH coupled with PFDM. In section \textbf{III}, we present the modified quantities of thermodynamics due to the thermal fluctuation. Also in section \textbf{IV}, we discuss the phase transition. In section \textbf{V}, we establish a relation between null geodesics and QNMs, and the last section summaries the results of our findings.
\section{The Space-time of PFDM around RN-AdS BH}
We consider the electromagnetic field and the PFDM (dark matter) field minimally coupled to gravity and cosmological constant as \cite{44}-\cite{46}
\begin{equation}
S=\int d^{4}x\sqrt{-g}\Big(\frac{1}{16 \pi G}R-\frac{1}{8\pi G}\Lambda+\frac{1}{4}F^{\mu \nu}F_{\mu \nu}+ \mathcal{L}_{DM}\Big),
\end{equation}
were $G$ is the Newton gravitational constant, $\Lambda$ is the cosmological constant, $F_{\mu \nu}$ Faraday tensor of the electromagnetic field and $\mathcal{L}_{DM}$ is the DM Lagrangian density. By variation of above action, we get the following field equations
\begin{equation}
R_{\mu \nu}-\frac{1}{2}g_{\mu \nu} R+ \Lambda g_{\mu \nu}= (T_{\mu \nu }(DM)+T^{(em)}_{\mu \nu }+\overline{T}_{\mu \nu})8 \pi G=-8 \pi G T_{\mu \nu}.
\end{equation}
\begin{equation}
F^{\mu \nu}_{;\nu}=0,\,\,\,\ F^{\mu \nu ;\alpha}+F^{\nu \alpha ;\mu}+F^{\alpha \mu ;\nu}=0.
\end{equation}
Here $T_{\mu \nu }$ is the energy-momentum tensor of the DM, $T^{(em)}_{\mu \nu }$ is the energy momentum tensor for the electromagnetic field and $\overline{T}_{\mu \nu}$ is the energy-momentum tensors of the matter. The Lagrangian density for electromagnetic field is $\mathcal{L}_{DM}=\frac{1}{4}F^{\mu \nu}F_{\mu \nu}$ and electromagnetic energy-momentum tensor is $ T^{(em)}_{\mu \nu }=\frac{1}{4\pi}(g^{\psi\omega}F_{\mu \psi}F_{\nu \omega}-\frac{1}{4 \pi} g_{\mu \nu}F_{\psi \omega}F^{\psi \omega})$. When BH is surrounded by the DM, it is considered that the DM must be perfect fluid. We can write the energy-momentum tensor in such a way that $T^{t}_{t}=-\sigma$, $T^{r}_{r}=T^{\theta}_{\theta}=T^{\phi}_{\phi}=\tau$. Furthermore, the simplest DM is considered as PFDM throughout this work and suppose that  $T^{r}_{r}=T^{\theta}_{\theta}=T^{\phi}_{\phi}=\tau=T^{t}_{t}(1-\eta)$, where $\eta$ is constant.

The BH solutions for the charged AdS space-time with PFDM \cite{44}, \cite{45} is given as
\begin{equation}
ds^{2}= -N(r)dt^{2}+ \frac{1}{N(r)}dr^{2}+ r^{2}(dr^{2}+\sin^{2}\theta d\phi^{2}),
\end{equation}
\\ where
\begin{equation}
N(r)= 1- \frac{2m}{r}+ \frac{Q^{2}}{r^{2}}-\frac{1}{3}\Lambda r^{2}+\frac{\alpha}{r}\ln[\frac{r}{\alpha}].
\end{equation}
Here $Q$ is the BH charge, $\alpha$ is the parameter of intensity of PFDM and $m$ is mass of BH. If $\alpha\rightarrow0$, the given BH reduces to RN AdS BH, furthermore, if the charge $Q=0$, it becomes Schwarzschild AdS BH.
\section{Thermal Fluctuation and Corrected Thermodynamical Quantities}
In this section, we formulate corrected thermodynamical quantities by the means of correction in the entropy of AdS BH surrounded by PFDM  \cite{47}. These corrected quantities comprises of the entropy, internal energy, pressure, Gibbs free energy, Helmohltz free energy, enthalpy and specific heat.
 The static and spherically symmetric line element of charged AdS BH surrounded by PFDM is as given in Eq. $(4)$.
There are three event horizons including the cosmological horizon, cauchy horizon and finally event horizon as shown in Fig.\textbf{1}.
\begin{figure}[ht!]
\centering
(a)\includegraphics[width=.4\linewidth, height=2in]{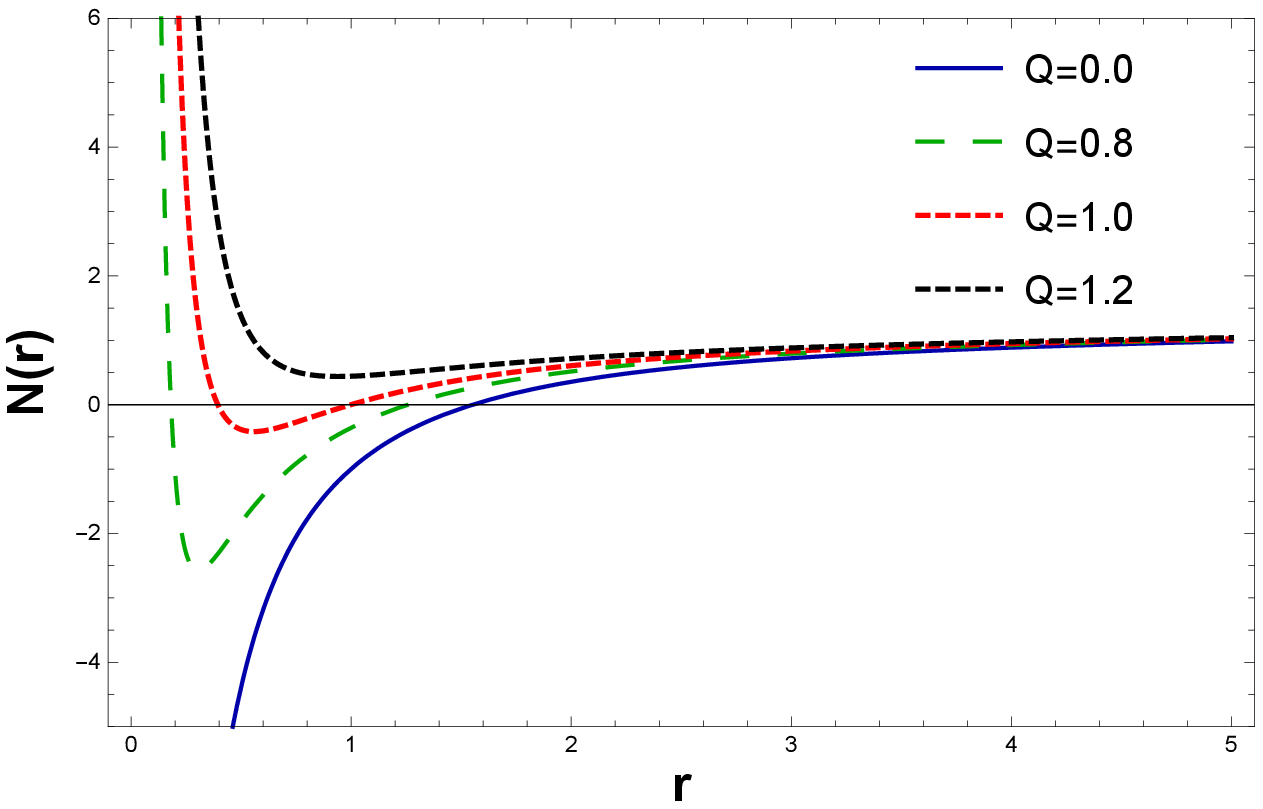}
(b)\includegraphics[width=.4\linewidth, height=2in]{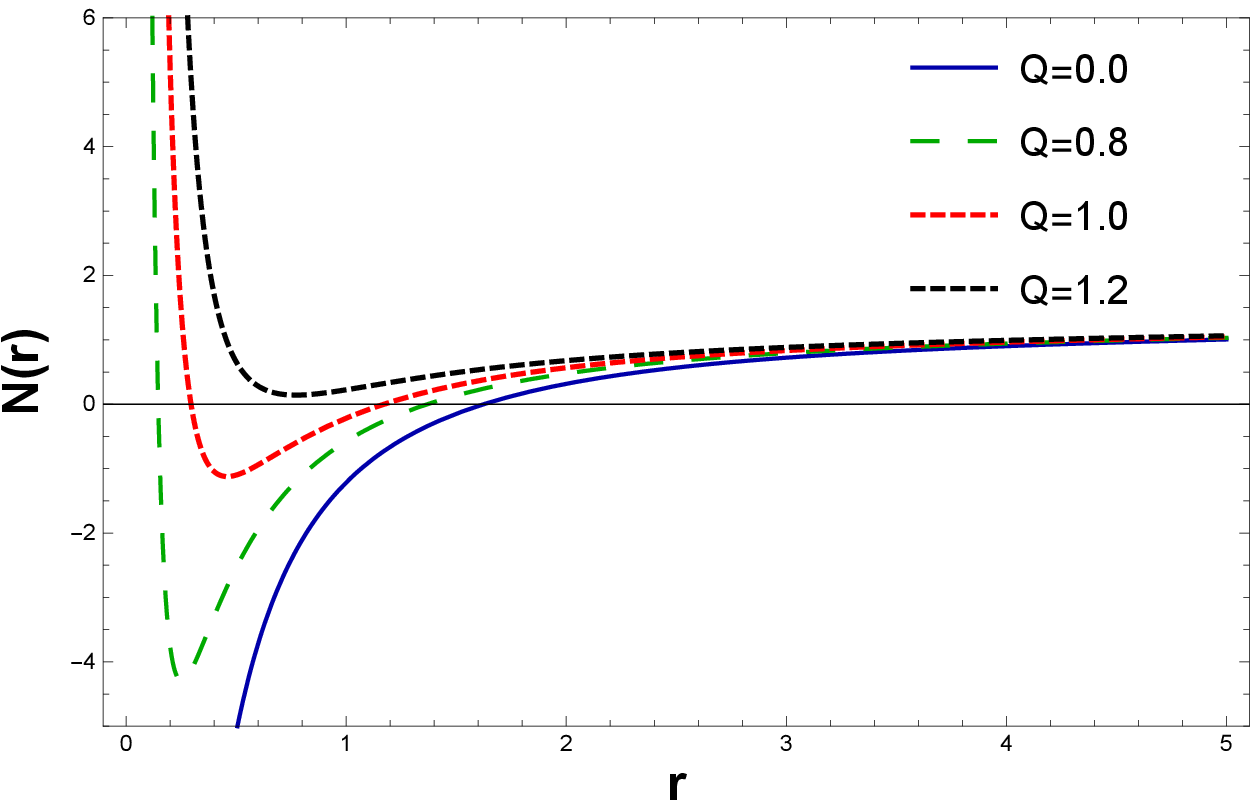}
(c)\includegraphics[width=.4\linewidth, height=2in]{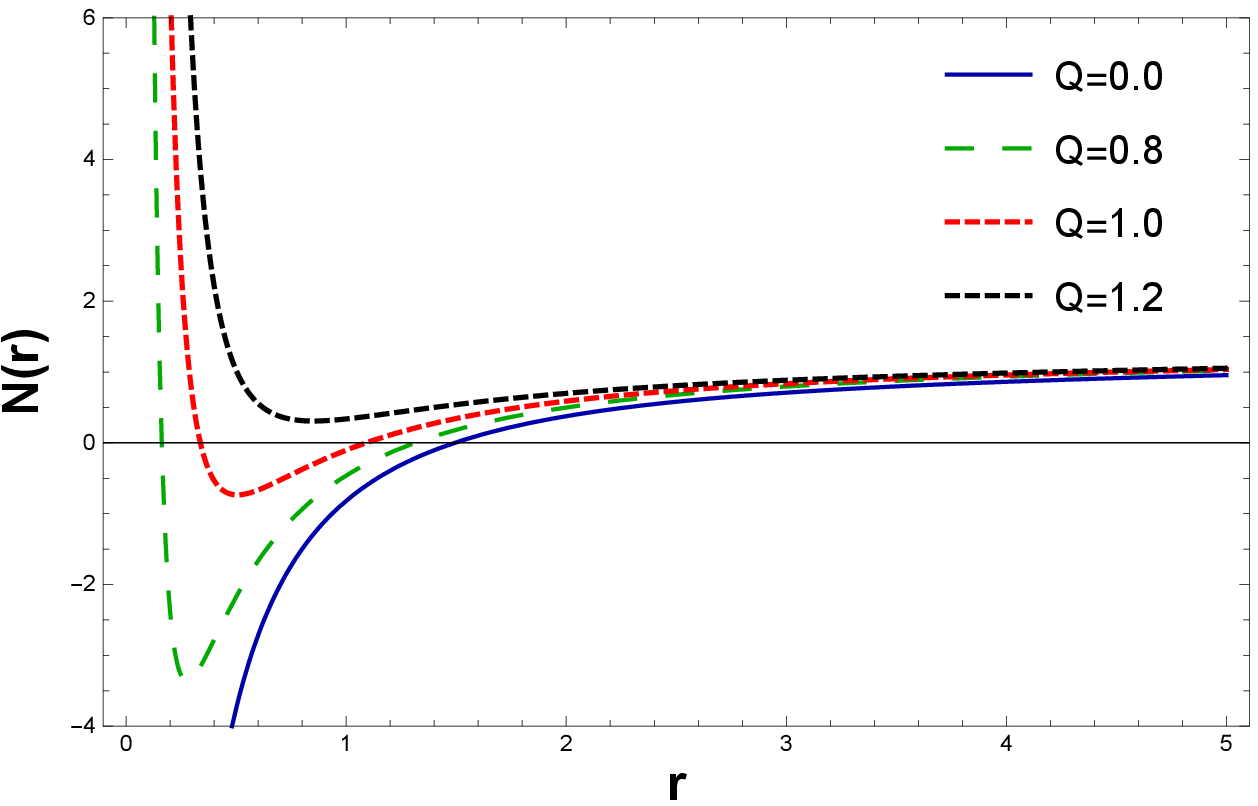}
\caption{The graph of Event Horizon as a function of $r$ for $l=20$ and $M=1$ and for (a) $\alpha=1$, (b) $\alpha=1.1$, and (c) $\alpha=1.2.$ }
\end{figure}\\
 The geometrical mass of the considered BH is given as
\begin{equation}
m=\frac{Q^2}{2r_{+}}+\frac{r_{+}}{2}+\frac{r_{+}^3}{2l^2}+\frac{\alpha}{2}\ln[\frac{r_{+}}{\alpha}].\label{a10}
\end{equation}
We can get the total mass of BH, by utilizing the definition of mass due to Abbott and Deser \cite{47a,47b}, as follows
\begin{equation}
M=\frac{m}{8}=\frac{1}{8}\Big(\frac{Q^2}{2r_{+}}+\frac{r_{+}}{2}+\frac{r_{+}^3}{2l^2}+\frac{\alpha}{2}\ln[\frac{r_{+}}{\alpha}]\Big).\label{a100}
\end{equation}

The Hawking temperature of given BH is,
\begin{equation}
T=\Big(\frac{3r_{+}}{4\pi l^2}-\frac{Q^2}{4\pi r_{+}^3}+\frac{\alpha}{4\pi r_{+}^2}+\frac{1}{4\pi r_{+}}\Big).\label{a11}
\end{equation}
The associated heat capacity ($ C=T \frac{ \partial S_{0}}{\partial T}$) can be written as
\begin{equation}
C=\frac{2 \pi  r_{+}^2 \left(l^2 \left(r_{+} (\alpha +r_{+})-Q^2\right)+3 r_{+}^4\right)}{l^2 \left(3 Q^2-r_{+} (2 \alpha +r_{+})\right)+3 r_{+}^4}.\label{a1100}
\end{equation}

\begin{figure}[ht!]
\centering
(a)\includegraphics[width=.4\linewidth, height=2in]{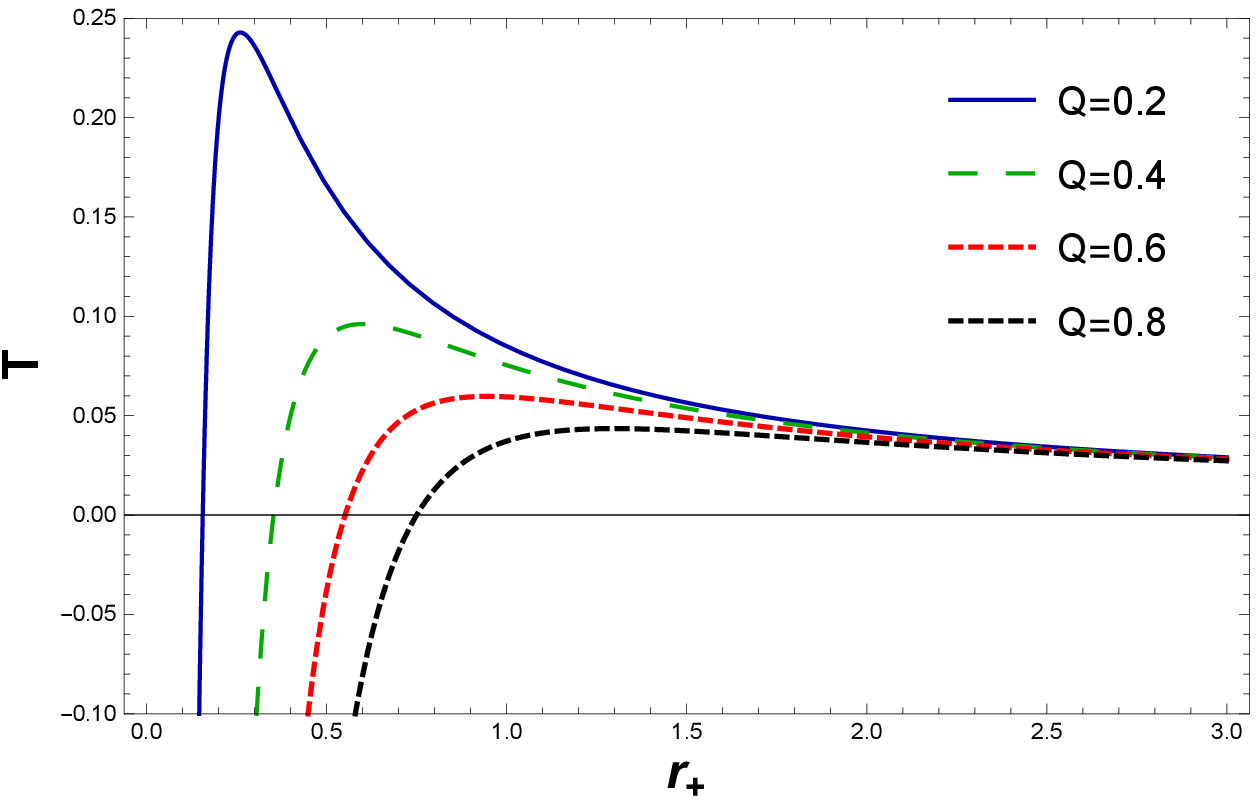}
(c)\includegraphics[width=.4\linewidth, height=2in]{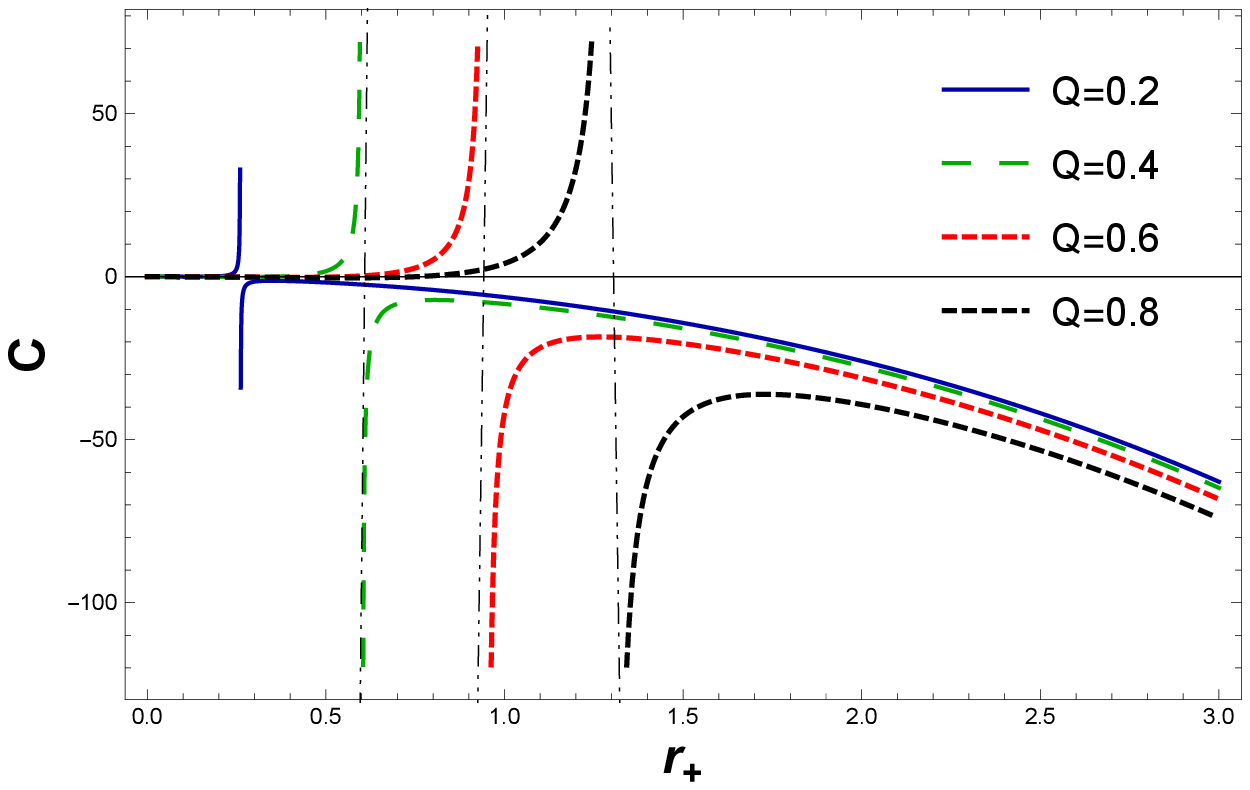}
(b)\includegraphics[width=.4\linewidth, height=2in]{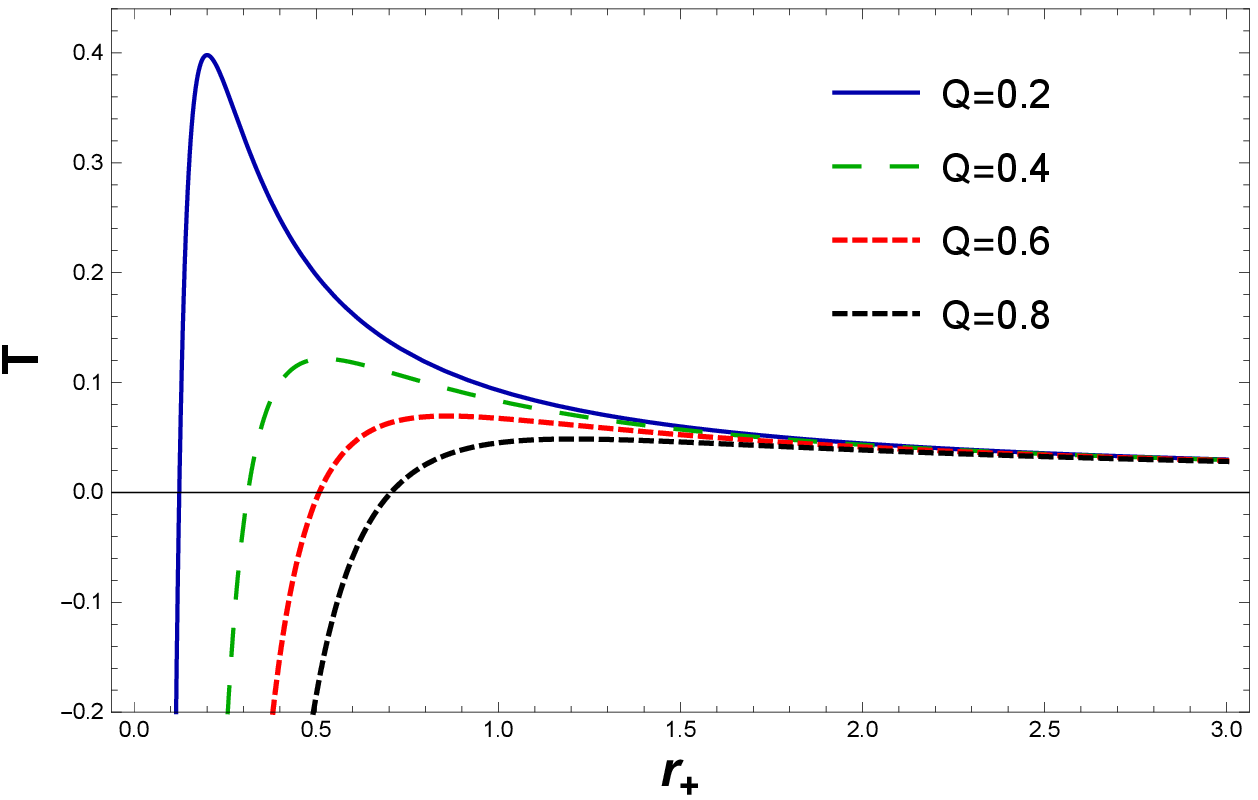}
(d)\includegraphics[width=.4\linewidth, height=2in]{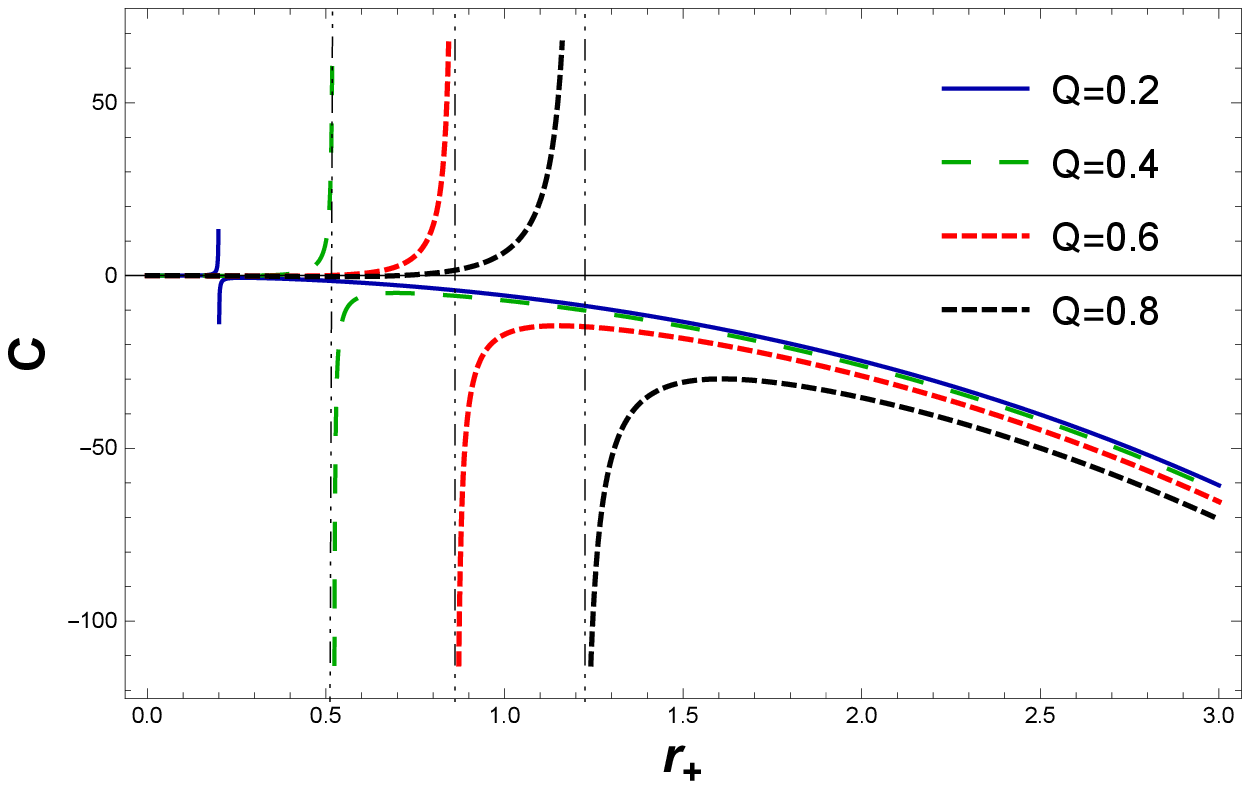}
\caption{In this fig  (letf 2 plots ) represent the graph of temperature as a function of $r_{+}$ for $l=20$. For (a) we take $\alpha=0.1$, for (b) $\alpha=0.2$. Similarly (right 2 plots ) represent the graph of heat capacity as a function of $r_{+}$ for $l=20$ and (c) we take  $\alpha=0.1$, for (d) $\alpha=0.2$ }
\end{figure}
In order to see the behavior of the temperature for the distinct values of the electric charge $Q$ and PFDM parameter $\alpha$, we have plotted the temperature in Fig. $2$ (left two plots). We observe that with smaller values of charge and fix value of $\alpha$, the temperature grows to maximum and then remain positive in given domain. Similarly, heat capacity is graphically depicted in Fig. $2$ (right two plots), in order to see the stability for the system. The specific heat has discontinuity at particular points and showing phase transition between stable and unstable phases for the considered BH.\\
The volume of the BH is
\begin{equation}
V=\frac{4\pi r_{+}^3}{3}.\label{a12}
\end{equation}
The BH entropy can be written as
\begin{equation}
S_0= \frac{A}{4}=\pi r_{+}^2,\label{a13}
\end{equation}
were $r_{+}$ is the radius of the BH event horizon.\\
Entropy is related to the event horizon of BH \cite{1,2}, which plays a significant role to seek the thermal properties of a system. The second law of thermodynamics requires that BH entropy must be greater than any other entity of comparable volume. Therefore, thermal equilibrium between BH and thermal radiations is unachievable, which requires logarithmic modifications to Bekenstein's entropy-area relation. The partition function \cite{48} is used to obtain the corrected entropy expression, which is defined as,
\begin{equation}
Z(\beta)= \int^{\infty}_{0} e^{-\beta E}\rho(E)dE,\label{a130}
\end{equation}
where $\rho(E)$ and $E$ denote the quantum density and average energy, respectively. We use inverse Laplace transformation on the partition function to get
\begin{equation}
\rho(E)= \frac{1}{2\pi i}\int^{i \infty +\beta_{0}}_{-i \infty +\beta_{0}} e^{-\beta E}Z(\beta)d\beta = \frac{1}{2\pi i}\int^{i \infty +\beta_{0}}_{-i \infty +\beta_{0}}e^{S_{0}(\beta)}d\beta, \label{a110}
 \end{equation}
 where $\beta$ must be positive and $S(\beta)=\ln Z(\beta)+\beta E$ is the modified entropy for the BH which depend on Hawking temperature. By applying the Taylor expansion, we get
 \begin{equation}
 S(\beta)=S_{0}+\frac{1}{2} (\beta-\beta_{0})^{2}\frac{\partial^{2}S(\beta)}{\partial\beta_{2}}|_{\beta=\beta_{0}}.\label{a111}
  \end{equation}
 The corrected entropy satisfy the relation $\frac{\partial S}{\partial\beta}=0$ and $\frac{\partial^{2} S}{\partial\beta^{2}}>0$ at the point of equilibrium. Now putting the values of Eqs.(\ref{a111}) in Eqs.(\ref{a110}), we get
 \begin{equation}
 \rho(E)= \frac{e^{S_0}}{2\pi i} \int e(^{\frac{1}{2}(\beta-\beta_{0})^{2}\frac{\partial^{2}S(\beta)}{\partial\beta^{2}}}) d\beta,
 \end{equation}
 which can be easily simplified (by following \cite{49}) as follows

\begin{equation}
 \rho(E)= \frac{1}{\sqrt{2\pi}}e^{S_{0}}((\frac{\partial^{2} S}{\partial\beta^{2}})|_{\beta=\beta_{0}})^{\frac{-1}{2}}.
 \end{equation}
 Thus, after simplification and neglecting higher order correction (as in \cite{490}), we get
\begin{equation}
S=S_{0}-\frac{1}{2}\ln(S_{0}T^{2})+\frac{\omega}{S_{0}},
\end{equation}
where $S_{0}$ be zeroth order entropy while, $\omega$ is the correction parameter for the higher order correction. It is possible to re-write the modified entropy of BH by substituting a very general parameter $\gamma$ for the factor $\frac{1}{2}$, which improves the corrected terms \cite{14, 15} as follows
\begin{equation}
S=S_{0}-\gamma\ln(S_{0}T^{2})+\frac{\omega}{S_{0}},\label{a112}
\end{equation}

 For distinct values of $\gamma$ and  $\omega$, we have following remarks:

I- When $\gamma$, $\omega$ $\rightarrow 0$, we get uncorrected entropy of the BH.

II- For $\gamma$, $\omega$ $\rightarrow 1$, we get the higher order correction terms.

III- When $\omega$ $\rightarrow 1$, $\gamma$ $\rightarrow 0$, we get the second order correction.

IV- As $\omega$ $\rightarrow 0$, $\gamma$ $\neq 0$, we usually obtain first order logarithmic correction.

In this paper, we consider the fourth case ($\omega$ $\rightarrow 0$, $\gamma$ $\neq 0$), in order to discuss the thermodynamics characteristics with the simple logarithmic corrections.
Using the  Eqs. (\ref{a11}) and (\ref{a13}) in Eq. (\ref{a112}), we can write
\begin{equation}
S= \gamma\log(16\pi l^4r_{+}^4)-2\gamma \log(l^2(r_{+}(\alpha+r_{+})-Q^2)+3r_{+}^4)+\pi r_{+}^2. \label{a113}
\end{equation}
\begin{figure}[ht!]
\centering
(a)\includegraphics[width=.4\linewidth, height=2in]{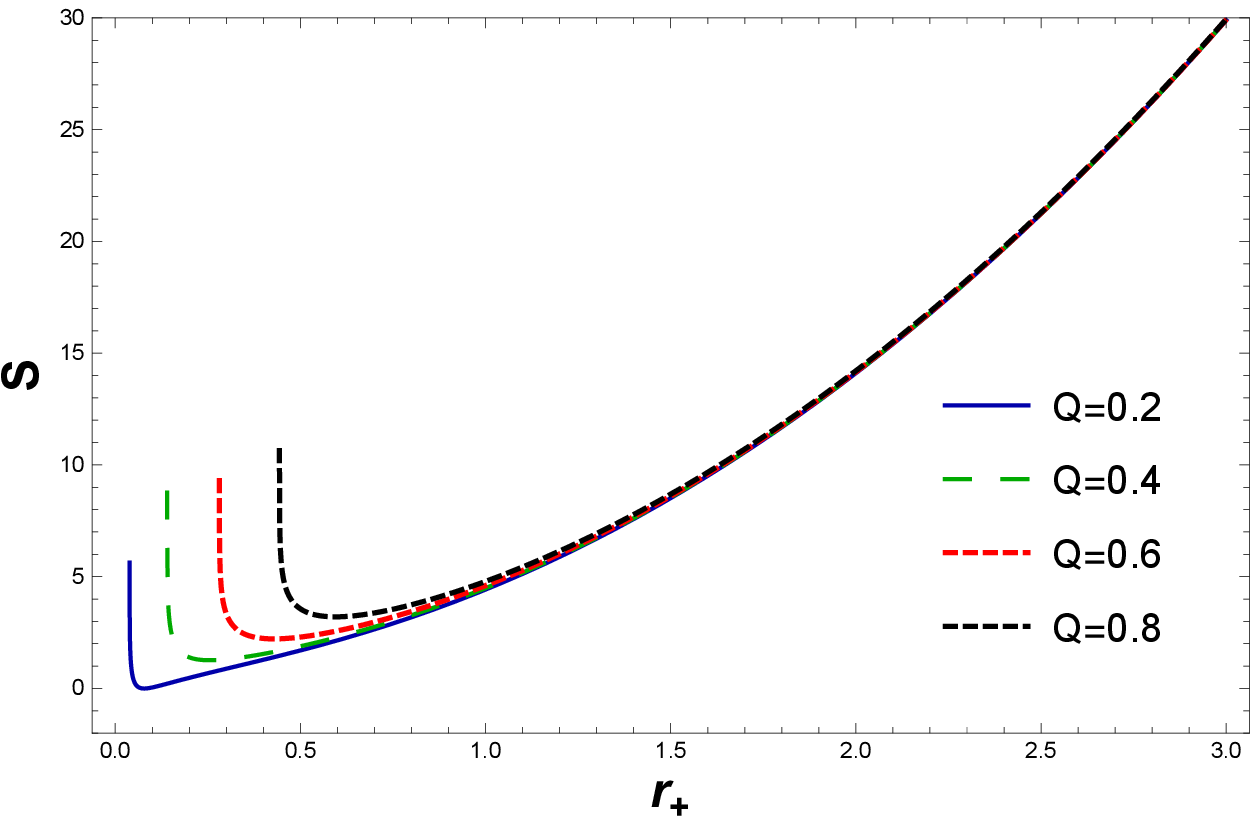}
(b)\includegraphics[width=.4\linewidth, height=2in]{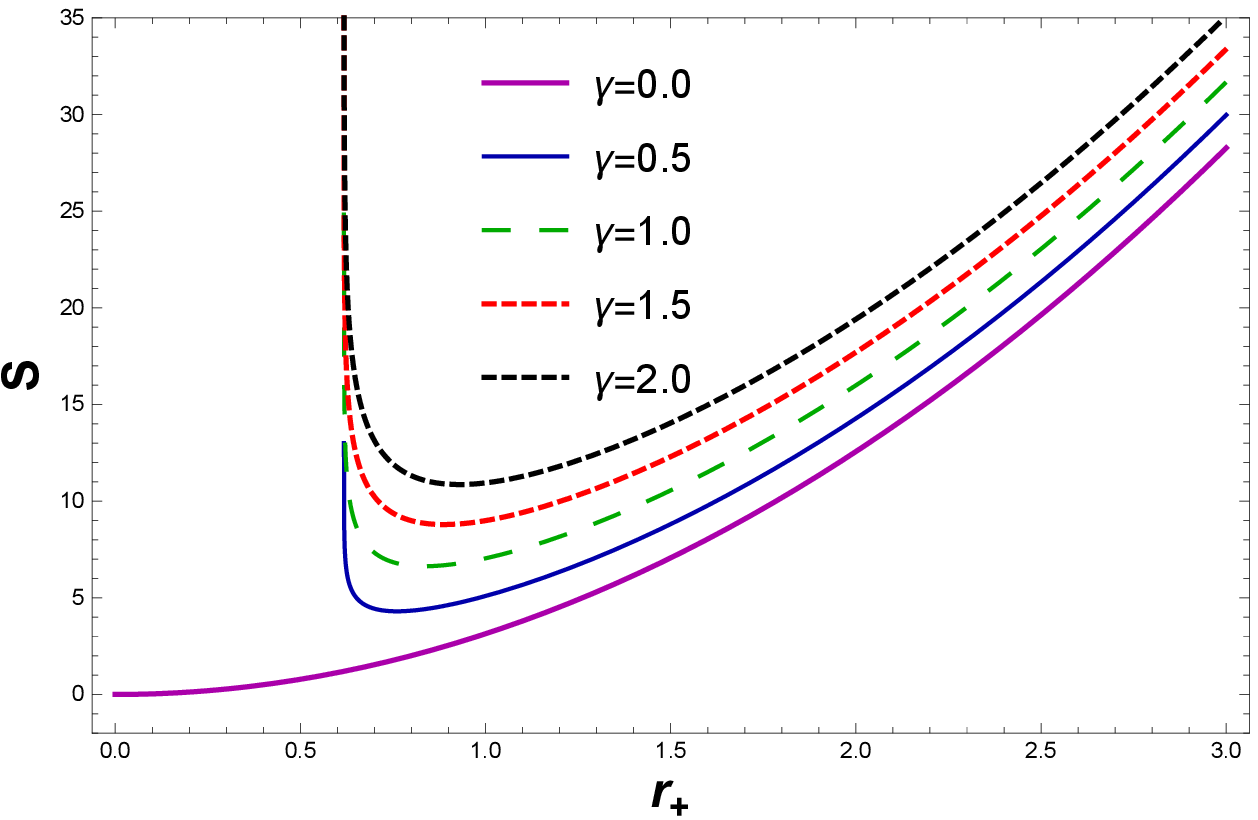}
(c)\includegraphics[width=.4\linewidth, height=2in]{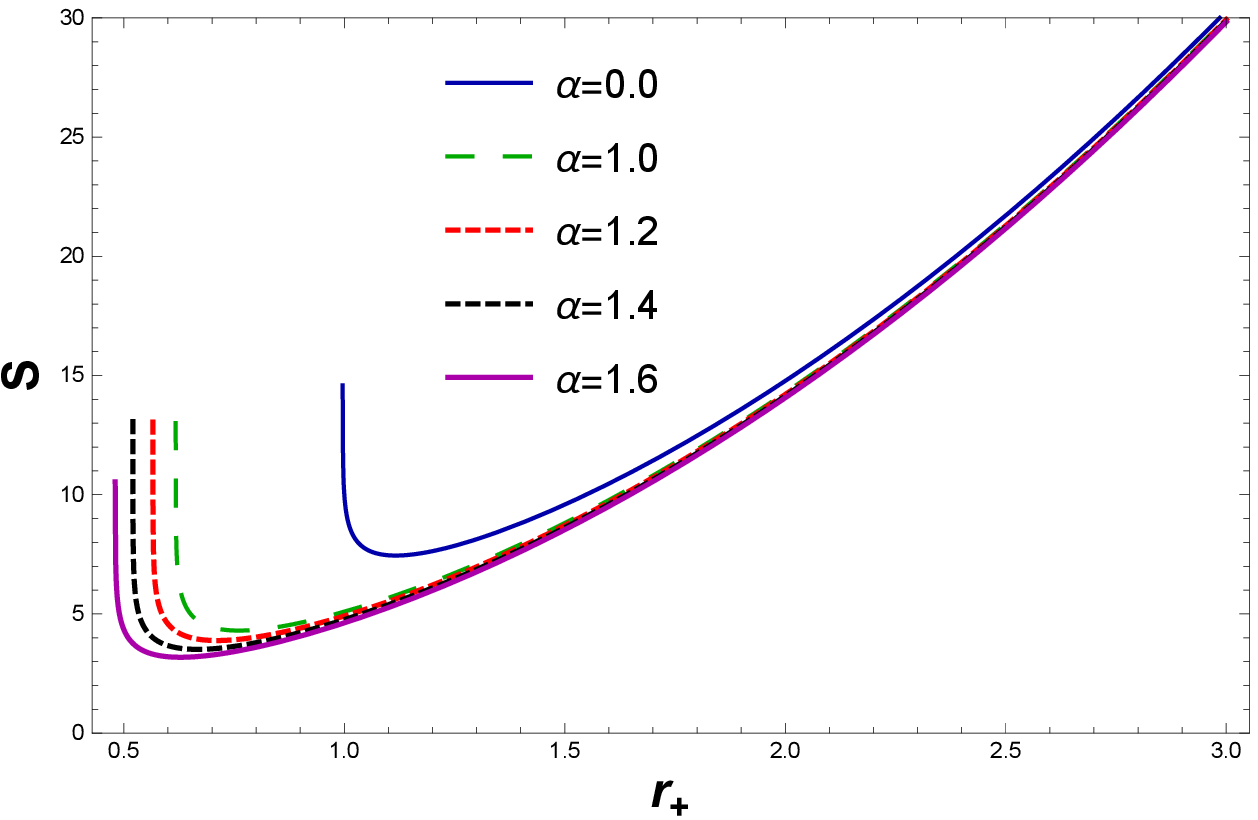}
\caption{This figure represent the graph of corrected entropy as a function of $r_{+}$ for $l=20$. For plot (a), we take  $\alpha=1$, $\gamma=0.5$, for (b) $Q=1$, $\alpha=1$  and for (c) $Q=1$, $\gamma=0.5$.}
\end{figure}\\
Figure $3$ shows that the corrected and uncorrected entropy is monotonically increasing smoothly for all parameters in the given domains. We can see that the given system for the large and small radius BHs sustains as positive and continues, and hence satisfies the second law of BH thermodynamics. Hence, similar results can be seen in \cite{14, 15} and \cite{490a}-\cite{490e}. They are all conclusively positive and have increasing trends, whether it is equilibrium entropy or corrected entropy. The usual entropy is only suitable for the small horizon radius BHs, while corrected entropy is more effective and sufficient for the large as well as small radius BHs. As thermal fluctuations are consequences of the statistical perturbations, it is clear that thermal fluctuation has a deep effect on small BHs, as shown in Fig.\textbf{3} plot \emph{(a)}, and plot\emph{(c)}, but has no effect, as shown in Fig.\textbf{3} plot \emph{(b)}, on small BHs. It is worthwhile to mention that plot \emph{(b)}, reveals the uncorrected and corrected entropy by means of $\gamma=0$, and altered values of the correction parameter. We examine that due to the thermal fluctuations, corrected entropy is always larger than usual entropy and deduce that the equilibrium (usual entropy) of BHs becomes unstable \cite{490f}. From plot\emph{(c)}, it is clear that $\alpha=0$ recovers the corrected entropy of RN-BH \cite{490g}, and distinct values of $\alpha$ explore the corrected entropy of RN-BH with the PFDM parameter, hence corrected entropy is deeply affected by small radii BHs. It may be noted that for very large BHs, thermodynamics is not affected by the small thermal fluctuation \cite{490a, 490c}.

In order to determine the the validity of first law of thermodynamics, which is most important, first we compute the total corrected mass, corrected Hawking temperature and corrected volume of the considered BH.
Now, we consider the thermodynamical equation of states under the modified entropy and Hawking temperature. The Helmohtz free energy can be modified as,
\begin{equation}
F=-\int S dT.\label{a01}
\end{equation}
Utilizing Eqs. (\ref{a11}) and (\ref{a113}) in Eqs. (\ref{a01}), we get the following from of Helmohtz free energy
\begin{eqnarray}\nonumber
F&&=\frac{1}{12\pi l^2r_{+}^3}(l^2(4\gamma Q^2+9\pi Q^2r_{+}^2+3\pi r_{+}^4-3\gamma\alpha r_{+})+6\pi\alpha l^2r_{+}^3\log(r_{+})+3\gamma(l^2(Q^2-r_{+}(\alpha+r_{+}))\\&&-3r_{+}^4)(\log(16\pi l^4r_{+}^4)-2\log(l^2(r_{+} (\alpha+r_{+})-Q^2)+3r_{+}^4))-3r_{+}^4(12\gamma+\pi r_{+}^2)). \label{a02}
\end{eqnarray}\\
\begin{figure}[ht!]
\centering
(a)\includegraphics[width=.4\linewidth, height=2in]{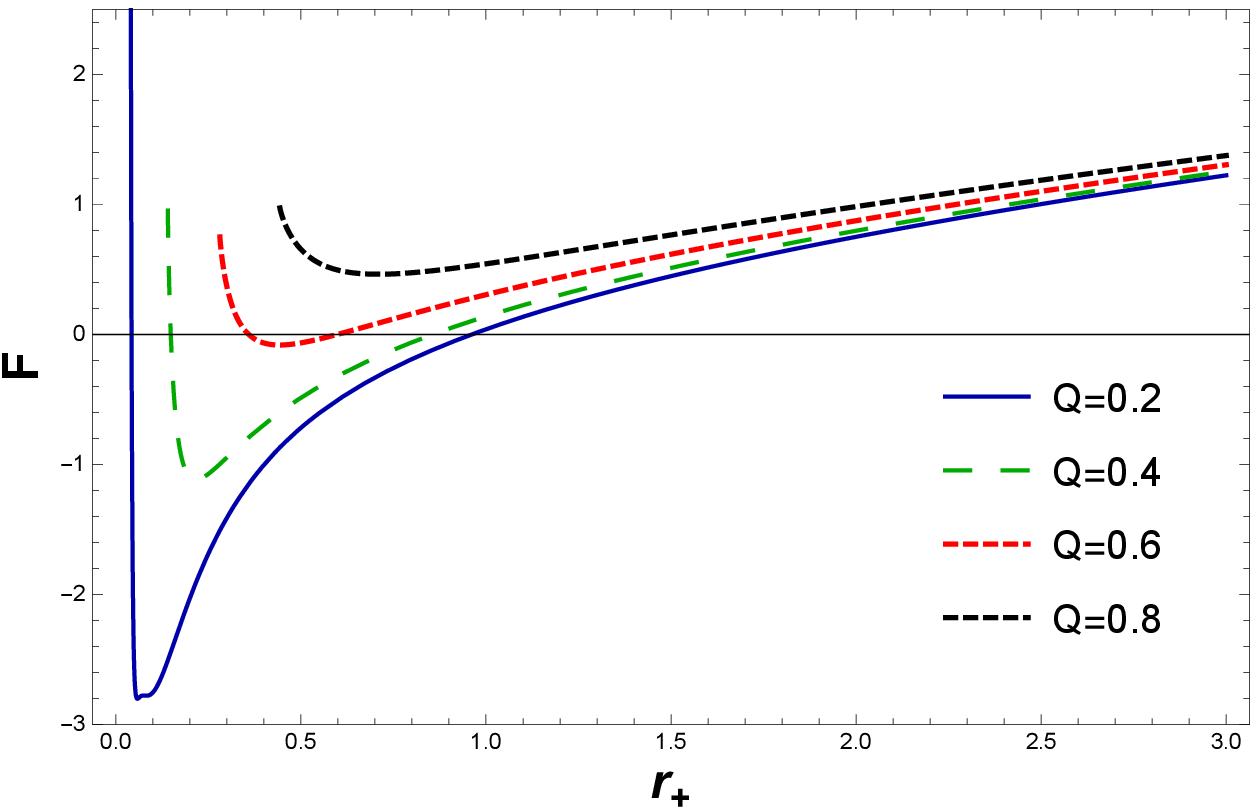}
(b)\includegraphics[width=.4\linewidth, height=2in]{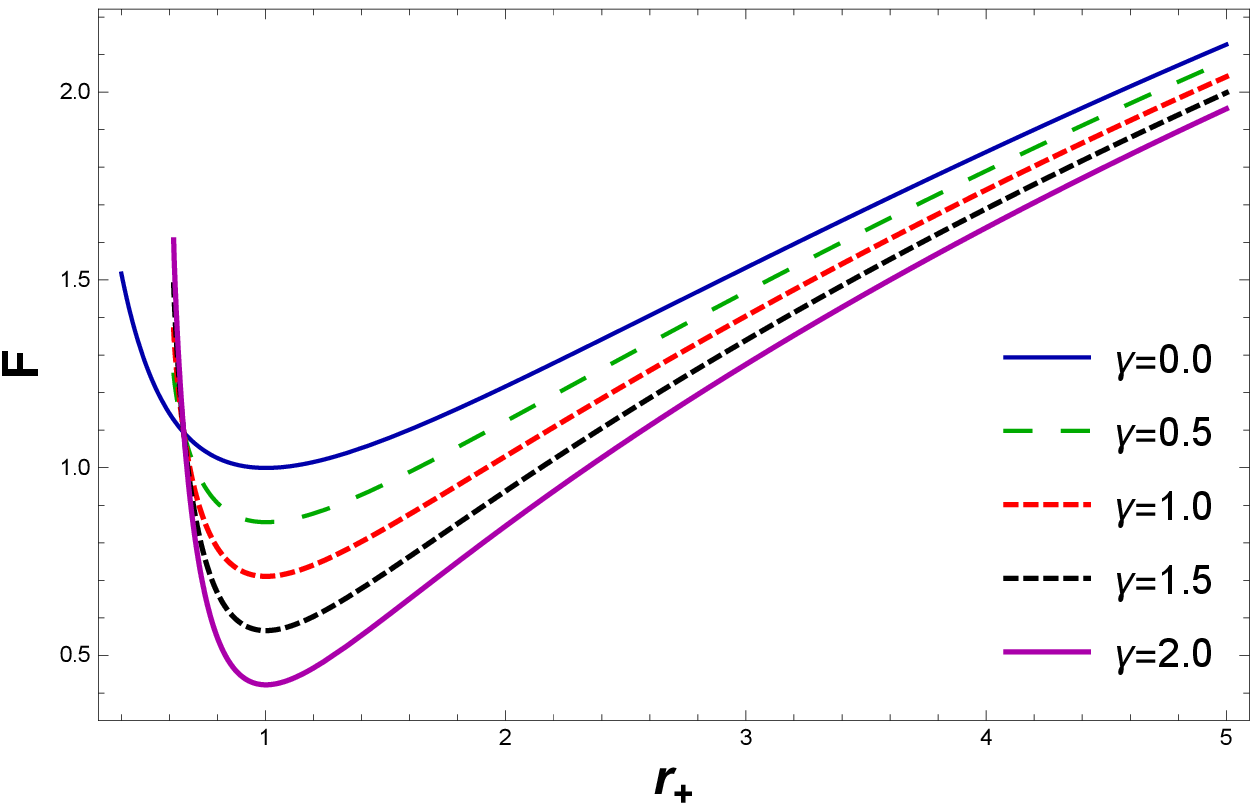}
(c)\includegraphics[width=.4\linewidth, height=2in]{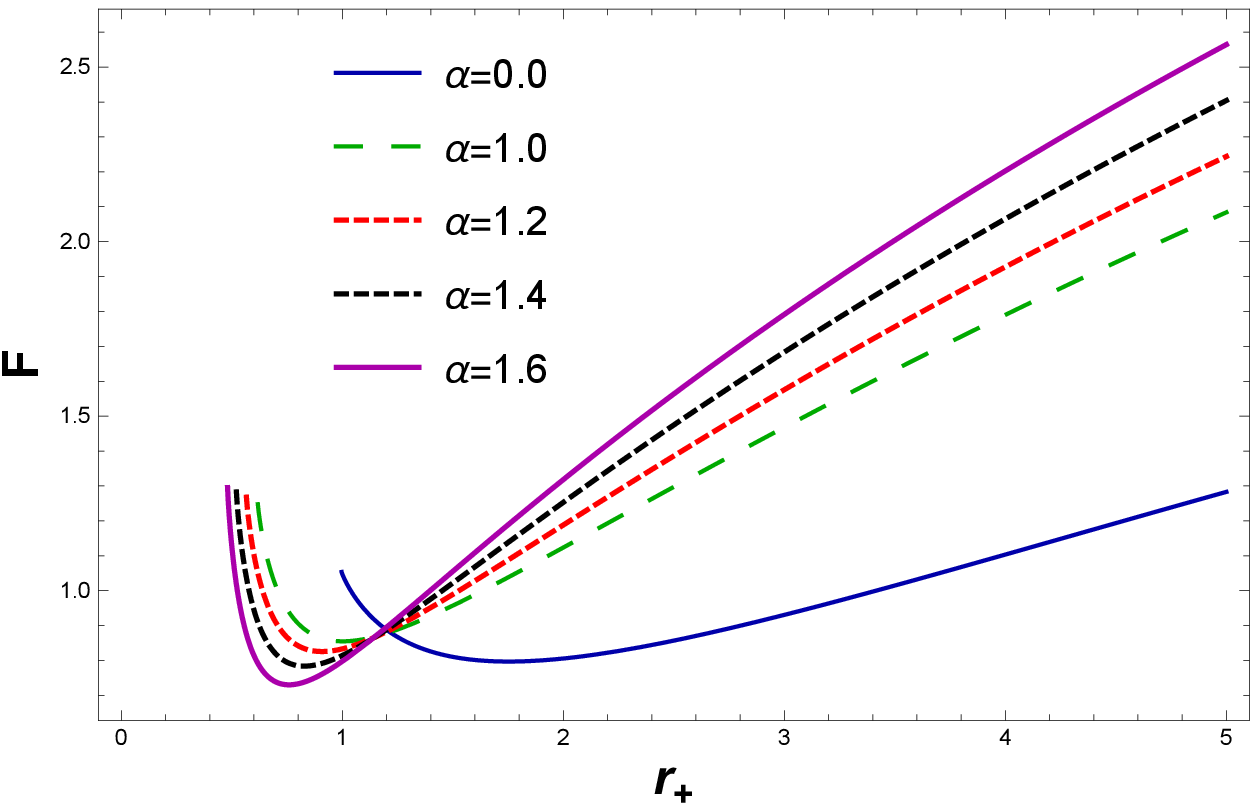}
\caption{The figure $4$ represent the graph of Helmohtz free energy as a function of $r_{+}$ for $l=20$ and (a) we take  $\alpha=1$, $\gamma=0.5$, for (b) $Q=1$, $\alpha=1$  and for (c) $Q=1$, $\gamma=0.5.$}
\end{figure}\\

Now Fig. $4$ shows the behavior of corrected Helmholtz free energy for the charged BH with PFDM. The Helmholtz free energy graphically indicate the increasing behavior for all parameters with increasing domains. The modifications of Helmholtz free energy can also be observed in some Rfs. \cite{14, 15, 490a, 490c, 490e, 490f, 490h, 490i}.
From Fig. \textbf{4} plot \emph{(a)}, it is observed that with the increasing values of the charge parameter, the Helmholtz free energy changes its phase from unstable to a stable system for the particular values of $Q=0.2$ to $Q=0.6$. For the higher values of the charged parameter $Q$ the Helmholtz free energy becomes stable for all considered domains as seen in \cite{15}.
 From Fig. \textbf{4} plot \emph{(b)}, we conclude that the Helmholtz free energy is increasing and positive function of the horizon radius without any correction. However, with the correction due to thermal fluctuations, the energy becomes less than the energy without correction for small as well as large BH radii. Hence, we conclude that logarithmic correction reduces the Helmholtz free energy rather than the energy with no correction, and charged AdS BH with PFDM becomes more stable in the presence of logarithmic correction. Similar results have been observed for decreasing the Helmholtz free energy due to the impact of fluctuations \cite{14, 490a, 490f}.
 Similarly, from Fig. \textbf{4} plot \emph{(c)}, we have seen graphically the modified Helmholtz free energy with and without the presence of PDFM parameter $\alpha$. The energy remains positive before and after the critical horizon radius. The critical horizon radius is a point where, there is no effect of thermal fluctuation on Helmholtz free energy. Finally, from the plot \emph{(c)}, one can conclude that initially, energy starts decreasing for the small BH horizon radius (before the critical radius) and increases for the large BH radii (after the critical radius). Moreover, in the presence of the PFDM parameter, energy becomes more stable and larger than energy without the PFDM parameter \cite{490g}.
Now, we are in position to determine the total corrected mass \cite{491}, we employ the concept of thermodynamical system along with following thermodynamical expression as,
\begin{equation}
\tilde{M}=F+ST, \label{a03}
\end{equation}
Using  Eqs. (\ref{a11}), (\ref{a113}), (\ref{a02}) in Eq. (\ref{a03}) , one can get the required corrected mass as,
\begin{equation}
\tilde{M}=\frac{1}{12}\Big( 3 \alpha -\frac{36 \gamma  r_+}{\pi  l^2}+\frac{6 r_+^3}{l^2}+\frac{4 \gamma  Q^2}{\pi  r_+^3}+\frac{6 Q^2}{r_+}-\frac{3 \alpha  \gamma }{\pi  r_+^2}+6 \alpha  \log \left(r_+\right)+6 r_+\Big).
\end{equation}
\begin{figure}[ht!]
\centering
(a)\includegraphics[width=.4\linewidth, height=2in]{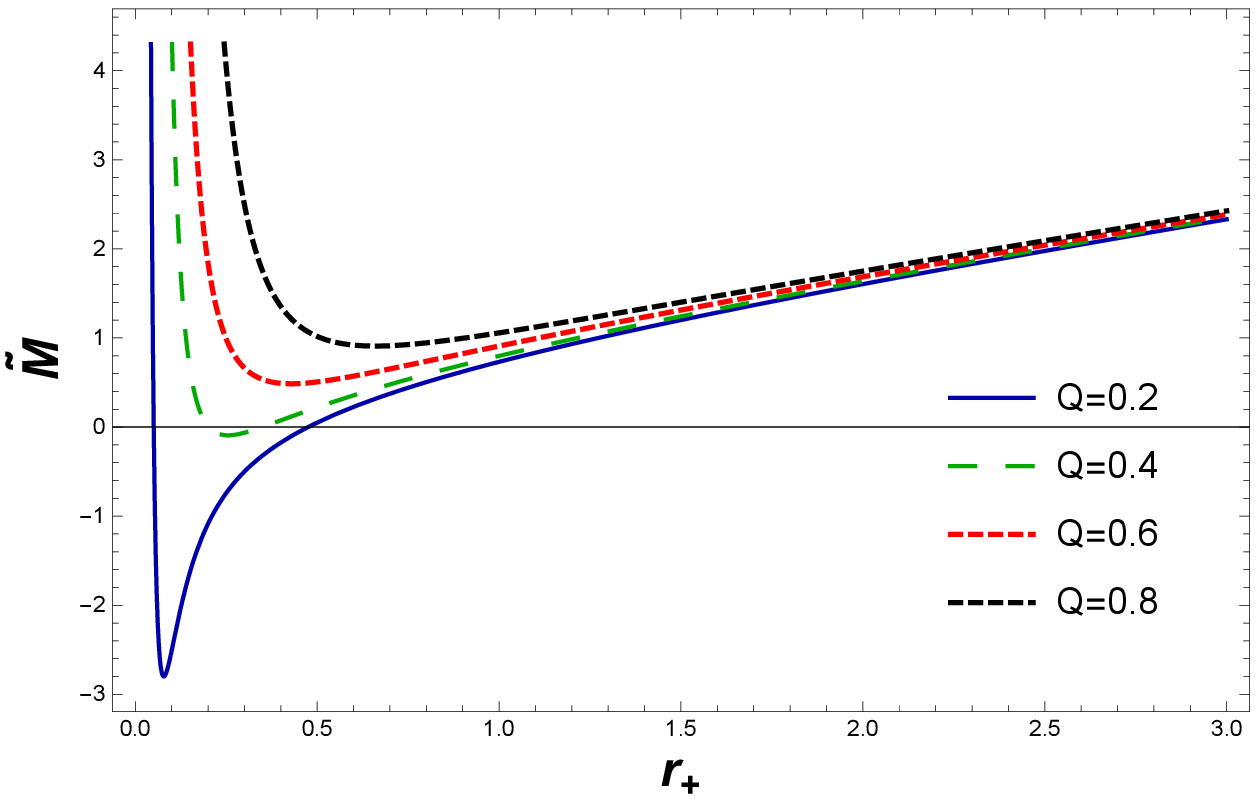}
(b)\includegraphics[width=.4\linewidth, height=2in]{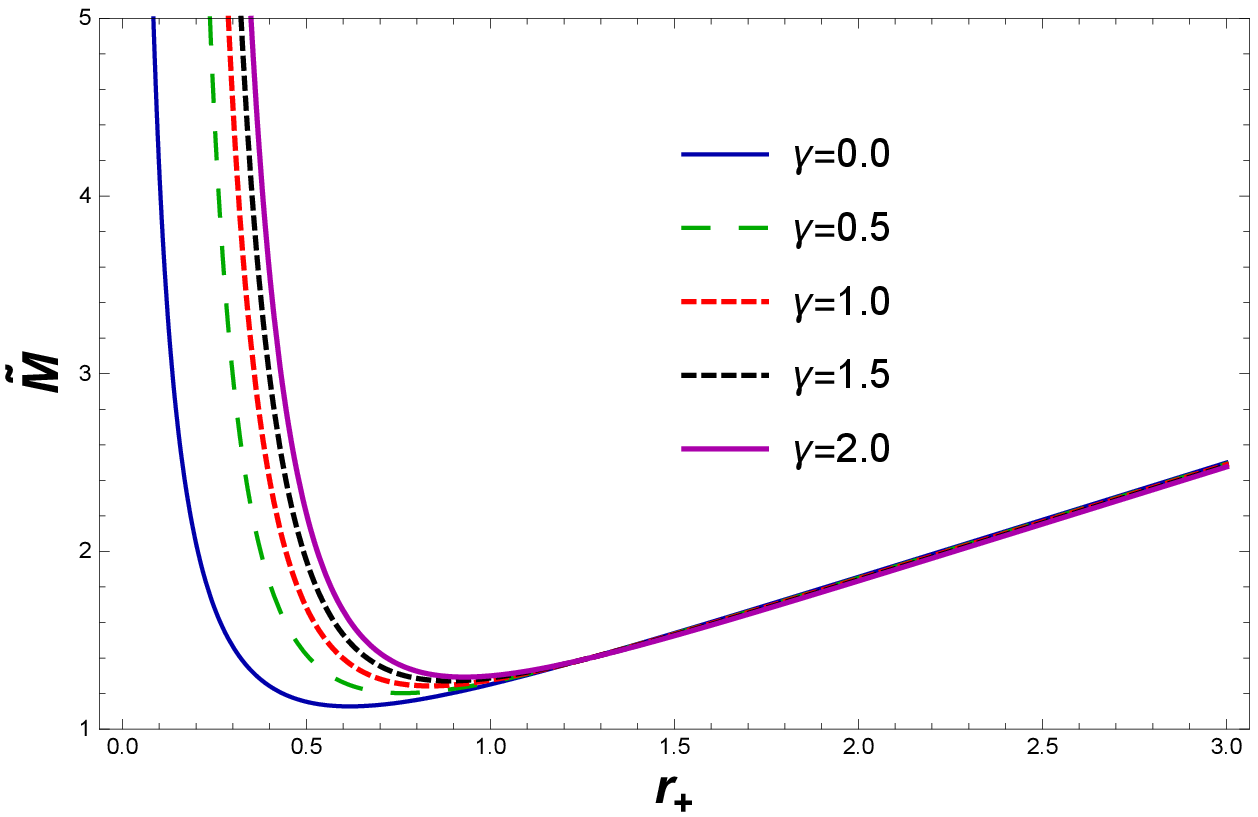}
(c)\includegraphics[width=.4\linewidth, height=2in]{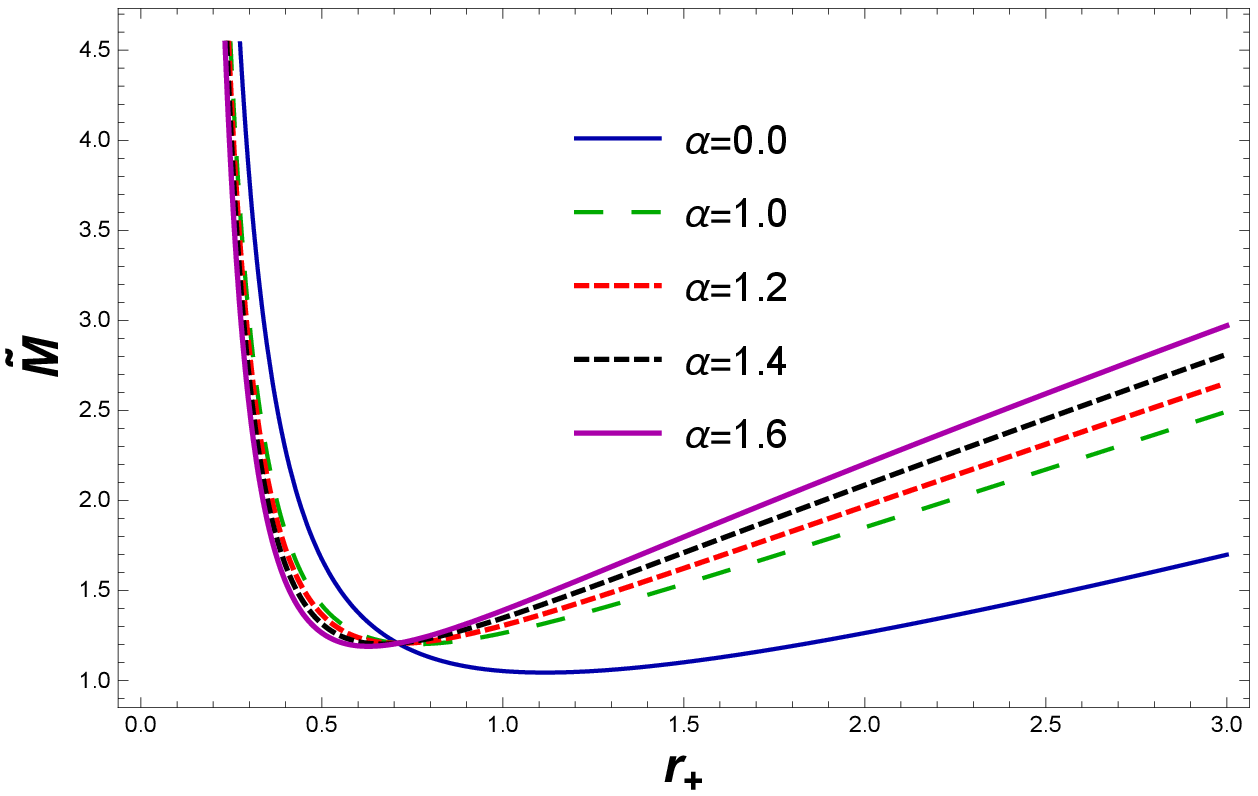}
\caption{In this Figs. represent the graph of total corrected mass as a function of $r_+$ for $l=20$. For (a) we take  $\alpha=1$, $\gamma=1$, for (b) $Q=1$, $\alpha=1$  and for (c) $Q=1$, $\gamma=0.5$.}
\end{figure}
 The behavior of the total corrected mass with respect to the horizon radius is depicted in Fig. $5$. It is worthwhile to note that the total corrected mass is stable throughout the given domain. Initially, mass decreases to the minimum and eventually increases to the maximum with the increasing values of all given parameters. Similar results are found in the literature \cite{490a, 490i, 491}, which conclude that the physical masses of the considered BHs evidently reach the maximum from the minimum level. From Fig.\textbf{5}, plot \emph{(a)}, it is easy to understand the physical behavior of the total corrected mass in terms of charged values.
 It is observed that with the increasing values of charge parameter $Q$, mass would divert from the negative to the positive region and remain stable for the higher values of the charge parameter. It is also important to mention that the occurrence of the phase transition is due to the charged parameters only for the horizon radius, and for the large horizon radius, the total internal energy of the system becomes larger. From Fig.\textbf{5}, plot \emph{(b)},we have depicted  the total physical mass regarding correction and without correction. Initially, the energy of the system will decrease to a minimum and eventually start increasing for large radii. The third plot in the Fig.\textbf{5}, shows the effects of PFDM parameter $ \alpha$ . When $ \alpha=0$, it recovers the total corrected mass for the RN-BH \cite{490g}, which reveals that there exists a critical horizon radius. It is found that the physical mass for both cases (with and without the PFDM) decreases before the critical point and increases (after the critical point) for large radii. Moreover, with the presence of PFDM, the internal energy of the system becomes higher.
It is important to mentioned that first law must hold for BH thermodynamical quantities under the thermal fluctuations, so, the modified first law of BH thermodynamics can be written as
\begin{equation}
\tilde{M}= \tilde{T}\delta S+\phi \delta Q+\tilde{V} \delta P=0,
\end{equation}
where $\tilde{T}$, $\phi$ and $\tilde{V}$ and denote the corrected temperature, electric potential, corrected and volume, respectively.
 Such modified thermodynamics characteristic can be sort out by following expressions as
\\
    $\tilde{T}= (\frac{\partial \tilde{M} }{\partial S})_{Q}$,\,\,\,\,\,\,\,\,\,\,\,\,\ $\phi=(\frac{\partial \tilde{M} }{\partial Q})_{S}$,\,\,\,\,\,\,\,\,\,\,\,\,\ $\tilde{V}=(\frac{\partial \tilde{M} }{\partial P})_{S, Q}$
   which gives
\begin{equation}
\tilde{T}=\Big(\frac{3 r_+}{4 \pi  l^2}-\frac{Q^2}{4 \pi  r_+^3}+\frac{\alpha }{4 \pi  r_+^2}+\frac{1}{4 \pi  r_+}\Big).
\end{equation}

\begin{multline}
\phi=\Big(\sqrt{\frac{2}{3}} \left(\frac{l^2 \left(2 \gamma +3 \pi  r_+^2\right)}{r_+}\right){}^{3/2} \left(-2 \gamma  l^2 Q^2-\pi  l^2 Q^2 r_+^2+\alpha  \gamma  l^2 r_++\pi  \alpha  l^2 r_+^3+r_+^4 \left(\pi  l^2-6 \gamma \right)+3 \pi  r_+^6\right) \\ \sqrt{\alpha  \gamma  l^2+r_+^2 \left(\pi  \alpha  \left(-l^2\right) \left(2 \log \left(r_+\right)+1\right)-2 r_+ \left(-6 \gamma +\pi  l^2+\pi  r_+^2\right)\right)}\Big)\\ \Big( \pi  l^4 r_+^2 (r_+^2 (\pi  \alpha  \gamma  l^2 (12 \log (r_+)+13)+r_+ (16 \gamma  (\pi  l^2-6 \gamma )+3 \pi  r_+ (\pi  \alpha  l^2 (2 \log (r_+)+3)\\+4 r_+ (\pi  l^2-4 \gamma )+8 \pi  r_+^3)))-2 \alpha  \gamma ^2 l^2)\Big).^{-1}
\end{multline}

\begin{multline}
\tilde{V}=\Big(8 r_+ \left(\pi  r_+^2-6 \gamma \right){}^2 \left(-2 \gamma  l^2 Q^2-\pi  l^2 Q^2 r_+^2+\alpha  \gamma  l^2 r_++\pi  \alpha  l^2 r_+^3+r_+^4 \left(\pi  l^2-6 \gamma \right)+3 \pi  r_+^6\right)\Big) \\\Big(3 l^2 (r_+ (18 \alpha  \gamma ^2+\pi  r_+ (r_+ (\pi  r_+(8 Q^2+r_+ (\alpha +6 \alpha  \log (r_+)+4 r_+))+\alpha  \gamma  (1-12 \log (r_+)))\\-16 \gamma  Q^2))-32 \gamma ^2 Q^2)\Big).^{-1}
\end{multline}

The corresponding pressure of the charged AdS BH with PFDM is given by
\begin{equation}
P= -\frac{dF}{dV},
\end{equation}
The pressure related to the cosmological constant is given by\\
The Eq. (52), given in Appendix.
\begin{figure}[ht!]
\centering
(a)\includegraphics[width=.4\linewidth, height=2in]{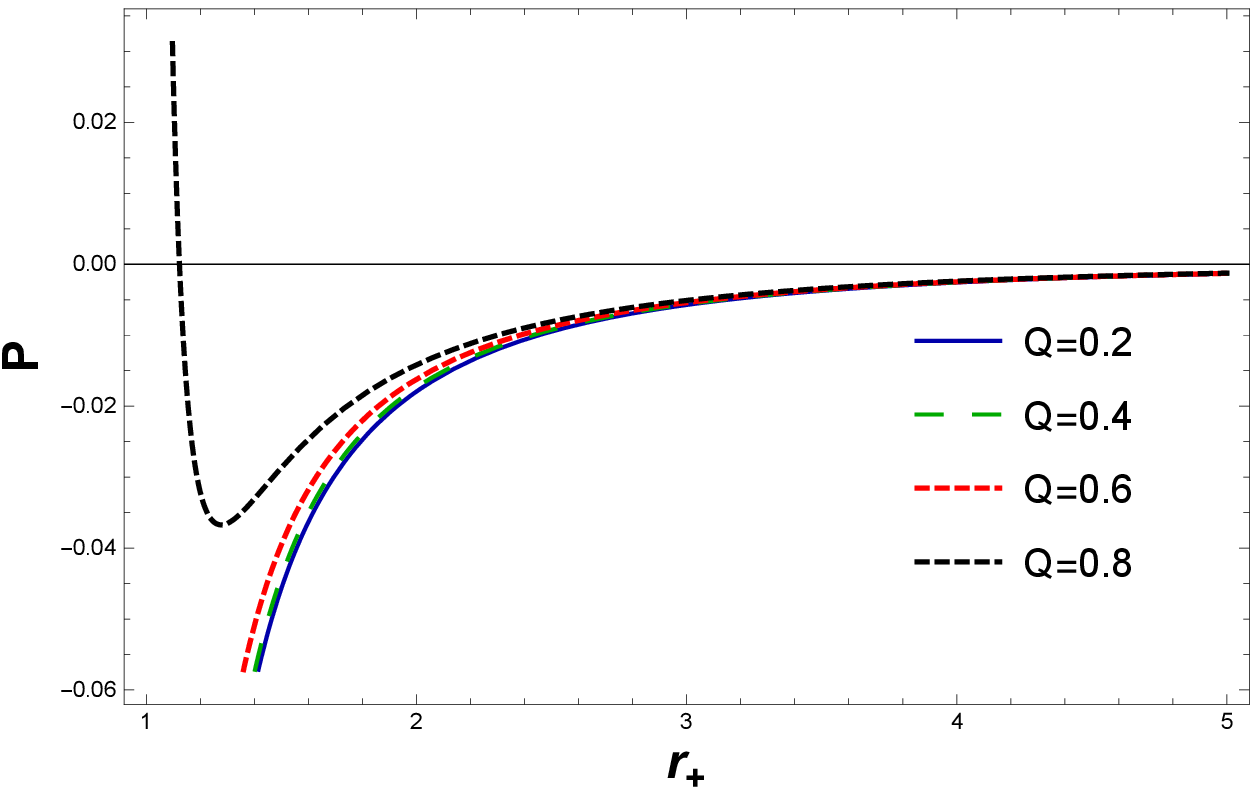}
(b)\includegraphics[width=.4\linewidth, height=2in]{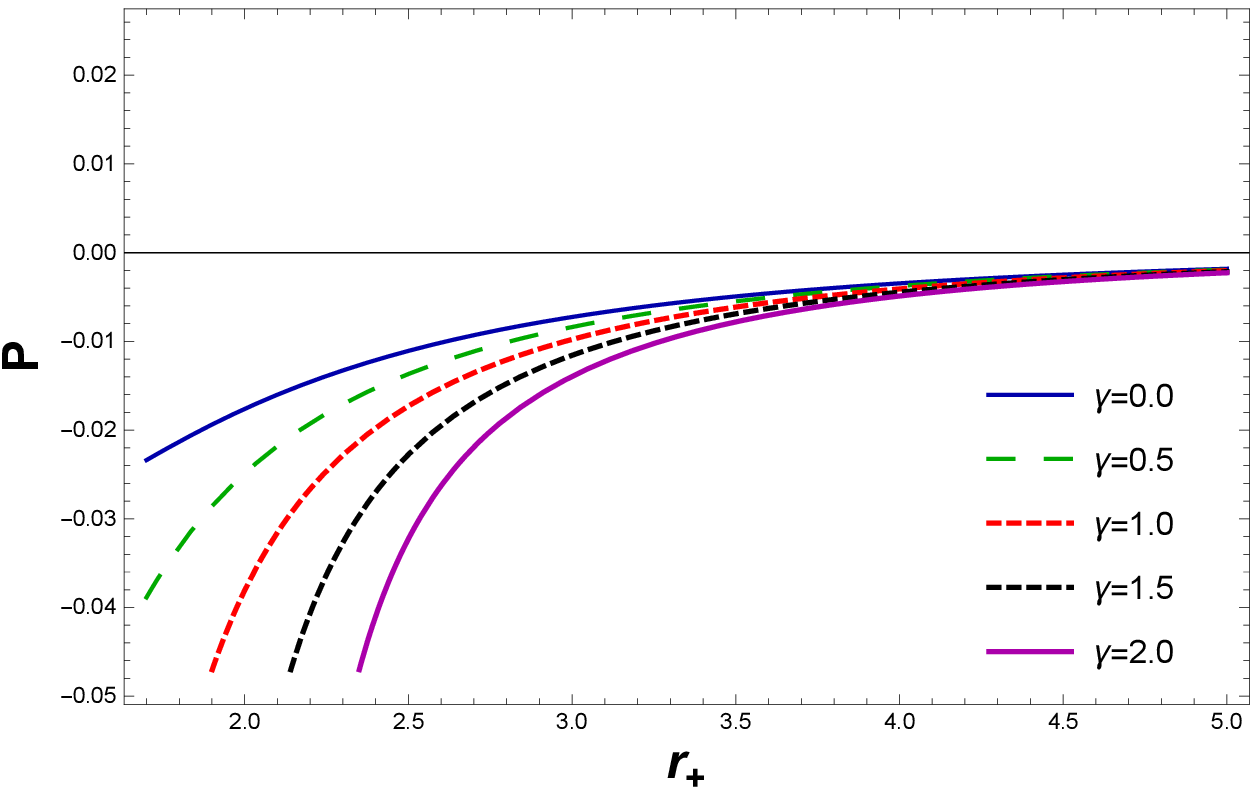}
(c)\includegraphics[width=.4\linewidth, height=2in]{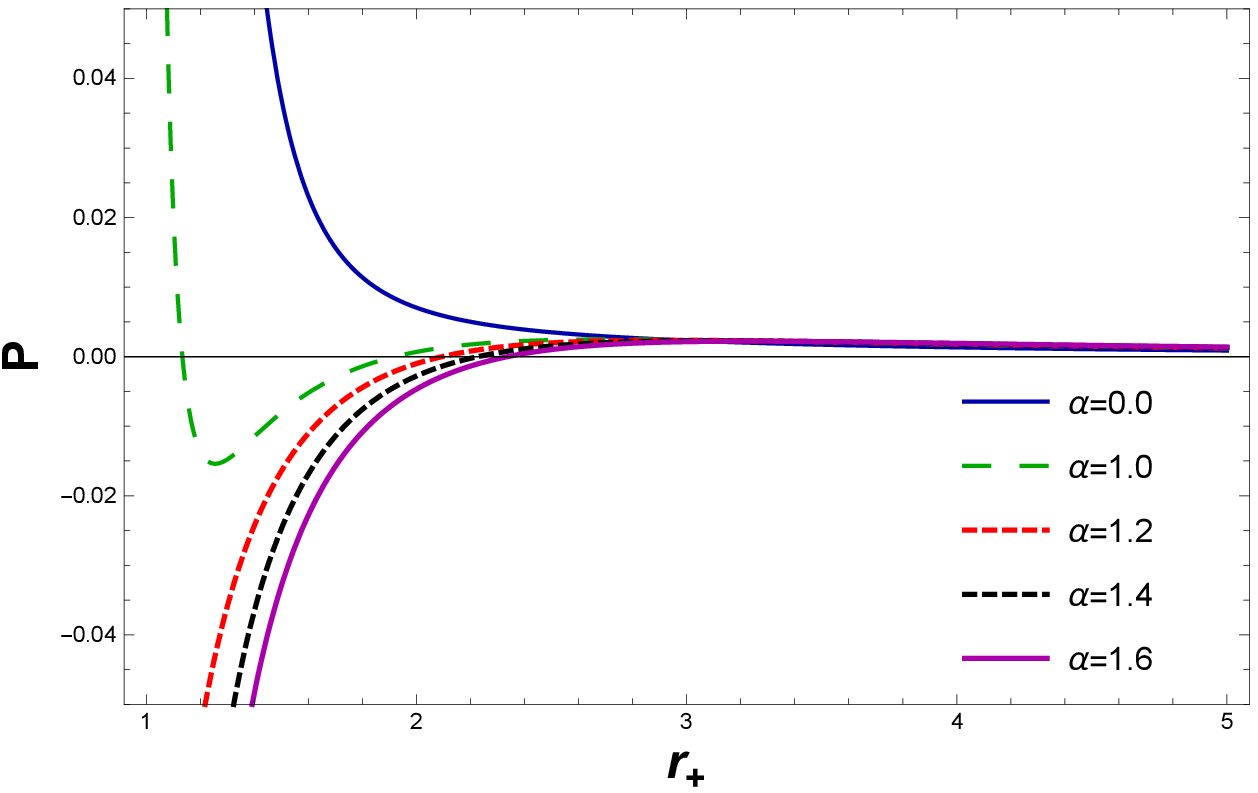}
\caption{In this fig represent the graph of pressure as a function of $r_{+}$ for $l=20$. For (a) we take  $\alpha=0.3$, $\gamma=0.5$, for (b) $Q=1$, $\alpha=1$  and for (c) $Q=1$, $\gamma=0.5$.}
\end{figure}
The modified pressure for the various Bhs have been observed in Refs.\cite{14, 15, 490f, 490c, 490j}. The authors in \cite{14, 15} have examined that the modified pressure decreased due to the presence of correction parameter. Thermal fluctuations are more effective for the small BTZ BHs \cite{490c}. Similarly, there is a remarkable impact of logarithmic correction on the pressure of the regular BHs \cite{490j}. We express the corrected pressure corresponding to the horizon radius $r_{+}$ in Fig. $6$. The plot \textit{(a)} starts decreasing initially at small values of charge parameter $Q$ and sustains in the negative region for the small size BHs. Noticeably, for the large BH (extremely large) horizon radius, the modified pressure (due to the influence of thermal fluctuations) of the BH rises and retains an equilibrium state (the stable state) for the fixed values of the PFDM parameter $\alpha$ and correction parameter $\gamma$.
With larger values of the charge parameter, the second order phase transition occurs from stability to instability and again to a stable system. The plot \textit{(b)}, compares the usual and corrected entropy parameters ($\gamma=0$, and $\gamma\neq0$), we see that the graph initially starts to increase from a negative to an equilibrium state. Hence, one can observe that the pressure and modified pressure remain negative for the small BH horizon radius. However, for the very large BH size, both (pressure and modified pressure) become stable under the impact of logarithmic correction terms. Also, it is observed that the uncorrected pressure is less negative than the corrected pressure for the AdS charged BH with PFDM. It is observed that as values of the PFDM parameter $\alpha$ increase for the small radius, the modified pressure changes its phase from unstable to stable region as shown in plot \textit{(c)}.

The enthalpy of the thermodynamical system can be computed as \cite{491}
\begin{equation}
H= \tilde{M}+PV,
\end{equation}
so, Eq. (53), given in Appendix.

In order to investigate the global stability, the Gibbs free energy is defined as
\begin{equation}
G= H-TS,
\end{equation}
From above mentioned quantities, the Gibbs free energy takes the form\\
The Eq. (54), given in Appendix.

\begin{figure}[ht!]
\centering
(a)\includegraphics[width=.4\linewidth, height=2in]{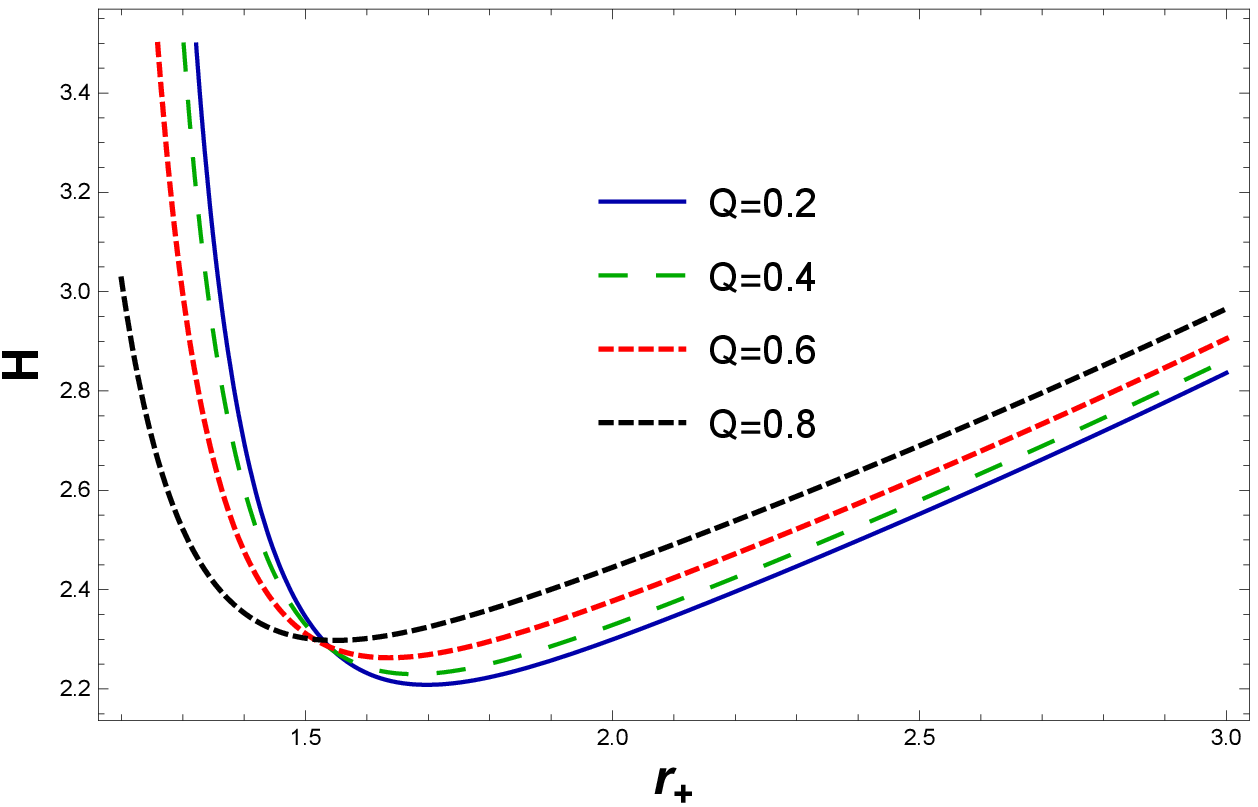}
(d)\includegraphics[width=.4\linewidth, height=2in]{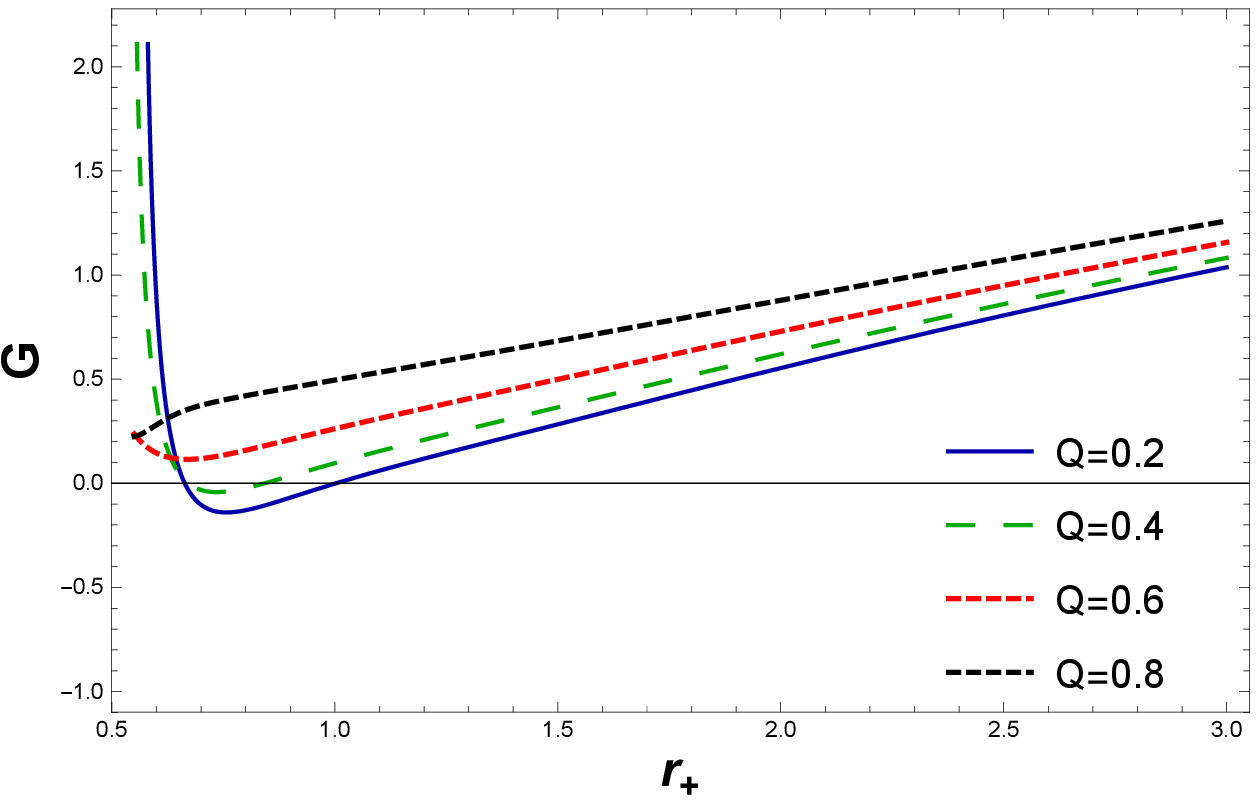}
(b)\includegraphics[width=.4\linewidth, height=2in]{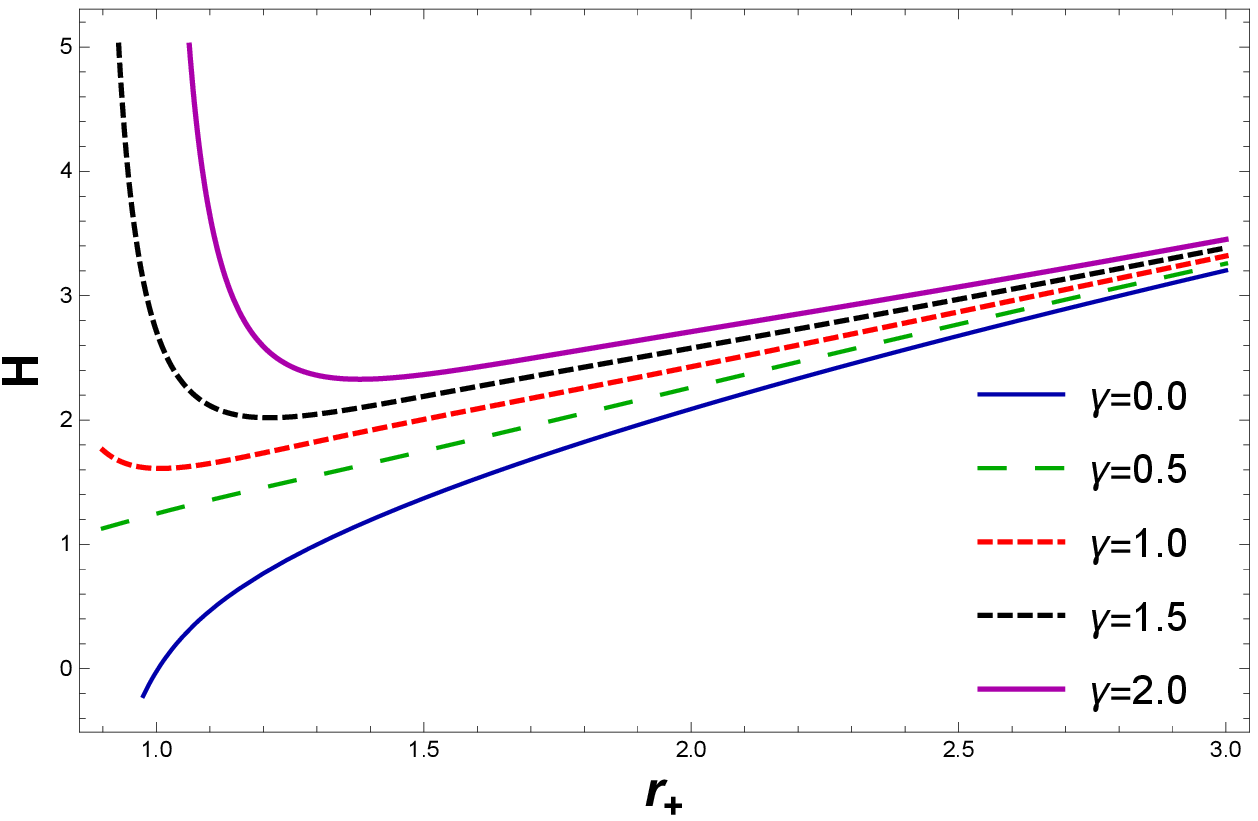}
(e)\includegraphics[width=.4\linewidth, height=2in]{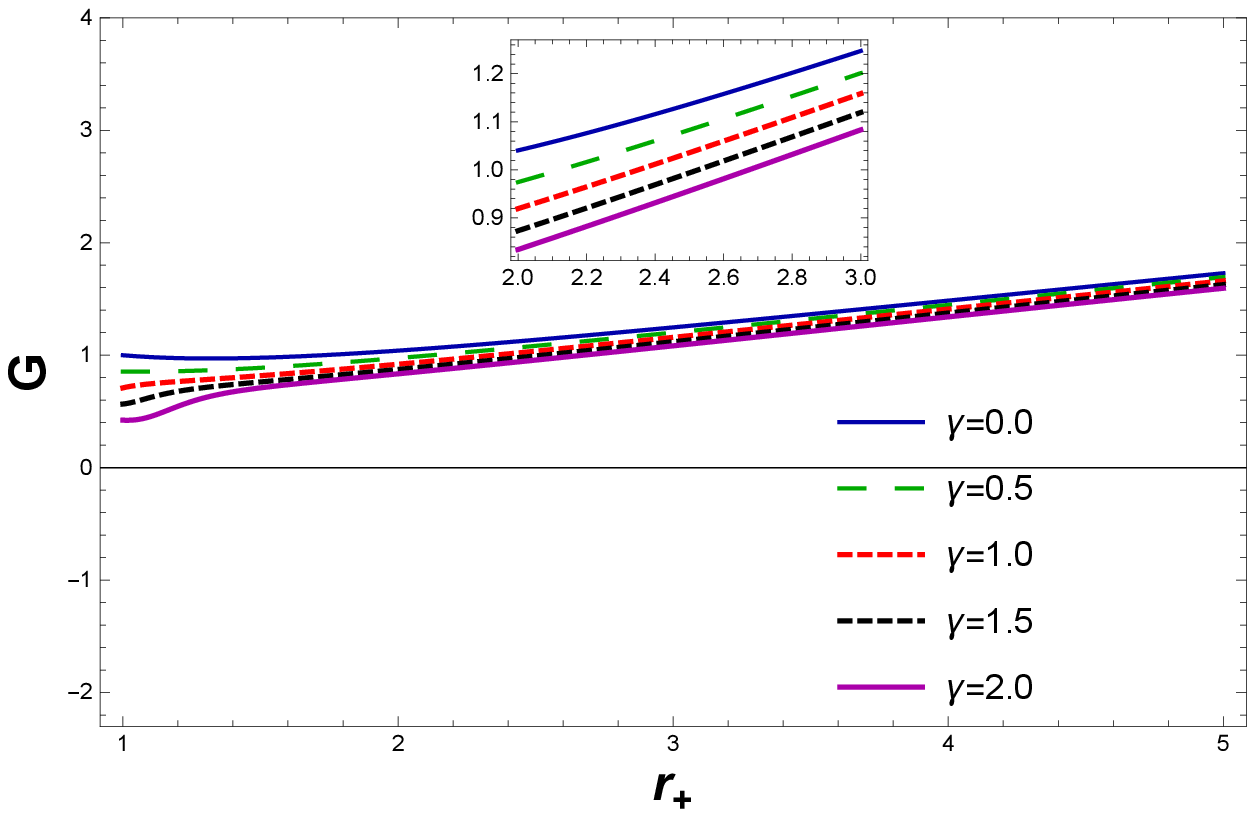}
(c)\includegraphics[width=.4\linewidth, height=2in]{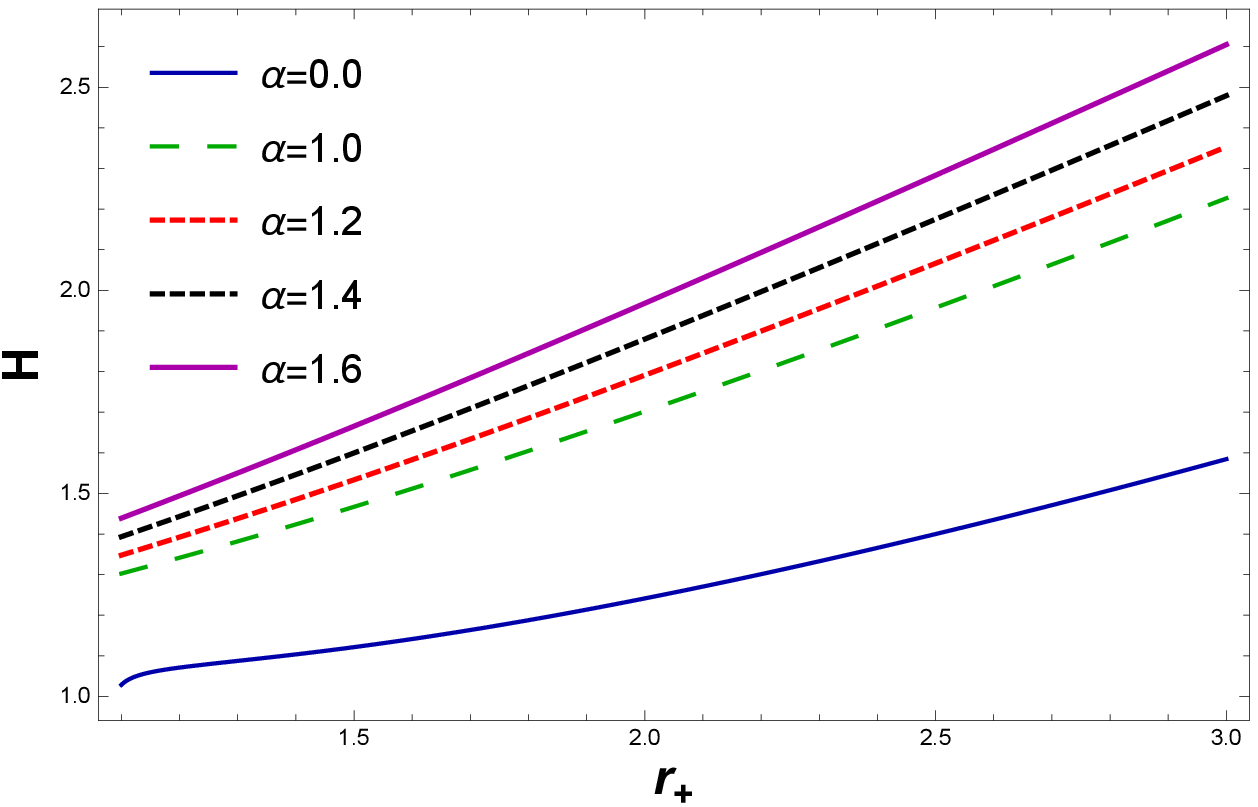}
(f)\includegraphics[width=.4\linewidth, height=2in]{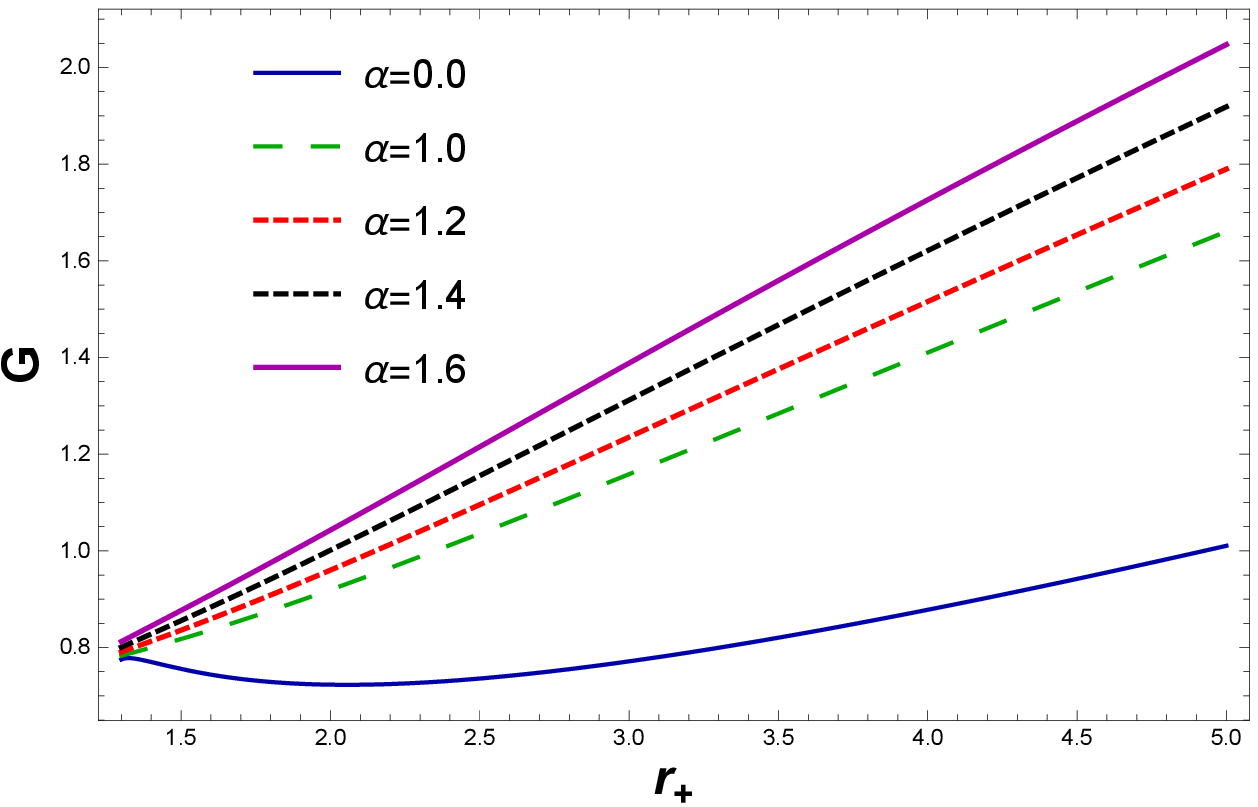}
\caption{In this fig  (letf 3 plots ) represent the graph of Enthalpy as a function of $r_{+}$ for $l=20$. For (a) we take  $\alpha=2$, $\gamma=3$, for (b) $Q=0.5$, $\alpha=3$  and for (c) $Q=1$, $\gamma=0.5$. Similarly (right 3 plots) represent the graph of Gibbs free energy as a function of $r_{+}$ for $l=20$. For (d) we take  $\alpha=1.5$, $\gamma=0.5$, for (e) $Q=1$, $\alpha=1$  and for (f) $Q=1$, $\gamma=1$}.
\end{figure}
Figure $7$ (left three plots), presents the graphical behavior of the corrected enthalpy with respect to the horizon, $r_{+}$. Over all, the enthalpy remains increasing and positive for the large BH radius, $r_{+}$. In the past, the enthalpy of the system (under the remarkable significance of thermal fluctuations) was also studied for various BHs in \cite{14, 29, 30, 490c, 490f}. The similar behavior were observed for the modified enthalpy as well as usual enthalpy in literature \cite{29}. In Ref. \cite{30}, enthalpy of the system in the presence of the correction parameter is more suitable for large radius BHs. As the values of the correction parameter decrease, the associated enthalpy starts increasing \cite{490c}. The enthalpy of the system decreases to a minimum for small horizon radii before the critical horizon radius, and after the critical point, the enthalpy increases gradually to its maximum value, which corresponds to the values of the charge parameter as shown in plot \textit{(a)}, giving stability for the large radii. A similar behavior can also be found in plot \textit{(b)}.
 It is clear that the enthalpy of the system under the influence of the correction parameter is higher than usual enthalpy. Which means that uncorrected enthalpy is unstable. Furthermore, in plot \textit{(c)}, with a fixed value of $Q$ and corrected parameter $\gamma$ the graph shows increasing behavior for the given radii. It is noted that enthalpy imply more stability in the presence of the PFDM parameter. Figure $7$, (right 3 plots) shows the Gibbs free energy versus horizon radius $r_{+}$ and demonstrates the physical significance of the Gibbs free energy with all three parameters. The graph of Gibbs free energy grows to maximum throughout the considered domain. Plenty of work is available in the literature \cite{14, 15, 490a, 490b, 490e} for the Gibbs free energy. The Gibbs free energy of the BH system decreases due to the involvement of thermal fluctuations, as observed by Pourhassan \cite{14}. Jawad \cite{15}, concludes that free energy would increase due to the large value of the cosmological constant. The physical interpretation of the Gibbs free energy in \cite{490a, 490b}, demonstrates that thermal fluctuation affect only small BHs. The Fig. $7$, plot \textit{(c)}, for the small values of charge parameter $Q=0.2$ to $Q=0.4$, Gibbs free energy changes its phase from negative region to positive. As the value of charge parameter $Q$ increases, the associated free energy becomes more stable under the effects of thermal fluctuations. It is interesting to note that phase transition occurs only for small horizon radii for the considered BH and free energy monotonically increases for large horizon radii. To check the comparison of equilibrium and modified energy, we see that the behavior of the plot \textit{(d)}, exhibits stability throughout the considered domain. There is no pattern difference between corrected and uncorrected free energy, but it is evidently clear that uncorrected Gibbs free energy is larger than the corrected one. In order to discuss the Gibbs free energy with the various values the PFDM parameter $\alpha$, we see the plot \textit{(f)}, it is clear that the free energy of the system becomes more stable and significant as the values of the charge parameter increase with $\alpha\neq0$ than the energy with $\alpha=0$.

Specific heat is yet another quantity that is used to study the local stability of the BH. The corrected heat capacity $ C_{s}=\frac{TdS}{dT}$ is given by
\begin{equation}
C_{s}= \frac{2(l^2(-2\gamma Q^2+\pi r_{+}^2(r_{+}(\alpha+r_{+})-Q^2)+\gamma\alpha r_{+})+3\pi r_{+}^6-6\gamma r_{+}^4)}{l^2(3 Q^2-r_{+}(2\alpha+r_{+}))+3 r_{+}^4}.
\end{equation}
The stability of the system can be checked by a test called the Hessian matrix of the Helmohtz free energy, which involves comparing the second-order derivatives of the Helmohtz free energy against the temperature and chemical potential $ \phi=\frac{\partial M}{\partial Q}$. The expression of the Hessian matrix \cite{50} has the form
\begin{equation}
H=\left(
\begin{array}{cc}
 H_{11} & H_{12} \\
 H_{21} & H_{22} \\
\end{array}
\right),
\end{equation}
where, $ H_{11}=\frac{\partial^{2}F}{\partial T^{2}}$,~~~$H_{12}=\frac{\partial^{2}F}{\partial T \partial\phi}$,~~~ $H_{21}=\frac{\partial^{2}F}{\partial\phi\partial T }$ ~~~~$ H_{22}=\frac{\partial^{2}F}{\partial \phi^{2}}$.
It is worthwhile to mention that the trace of the matrix $T_r(H)=H_{11}+H_{22}$ must be positive. Here
\begin{equation}
H_{11}=8\pi l^2r_{+}^3(\frac{\gamma}{l^2(r_{+}(\alpha+r_{+})-Q^2)+3r_{+}^4}+\frac{\gamma-\pi r_{+}^2}{l^2(3Q^2-r_{+}(2\alpha+r_{+}))+3 r_{+}^4}),
\end{equation}
\begin{multline}
H_{22}=-\frac{1}{4 \pi l^2Q^2r_{+}}(2\gamma(l^2(\alpha r_{+}-3Q^2)+3r_{+}^4)(\log(16\pi l^4r_{+}^4)-2\log(l^2(r_{+}(\alpha+r_{+})-Q^2)+3r_{+}^4))\\+\frac{1}{l^2(Q^2-r_{+}(\alpha+r_{+}))-3r_{+}^4})(2(l^4(\pi r_{+}^3(\alpha+r_{+})(r_{+}(\alpha +r_{+})-Q^2)+\gamma  (2Q^2-\alpha r_{+})(3Q^2-r_{+}(2\alpha+r_{+})))\\-3l^2r_{+}^4(\gamma(-8Q^2+2r_{+}^2+5\alpha r_{+})+\pi r_{+}^2(r_{+}(\alpha+r_{+})-2Q^2))+18r_{+}^8(\gamma-\pi r_{+}^2))).
\end{multline}
Thus the trace of Hessian matrix is given \cite{51} 
by Eq. (55) in Appendix.
\begin{figure}[ht!]
\centering
(a)\includegraphics[width=.4\linewidth, height=2in]{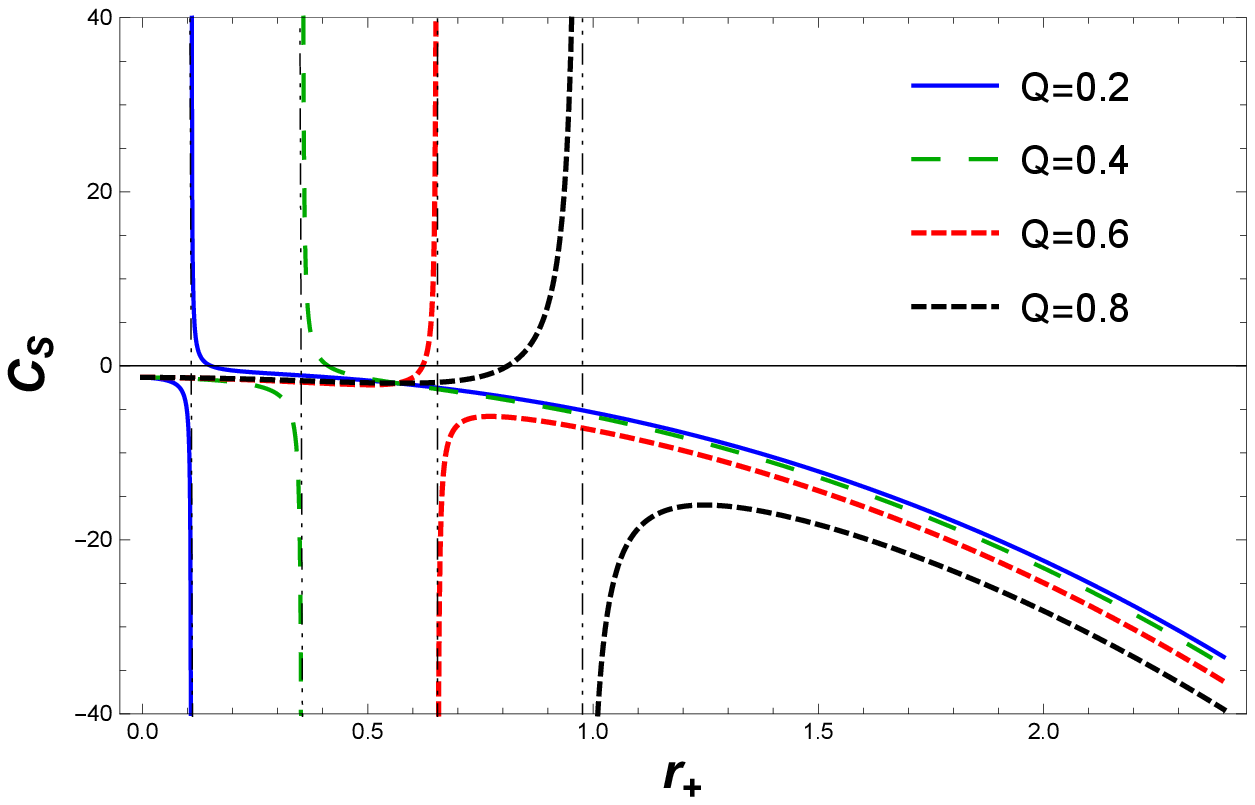}
(d)\includegraphics[width=.4\linewidth, height=2in]{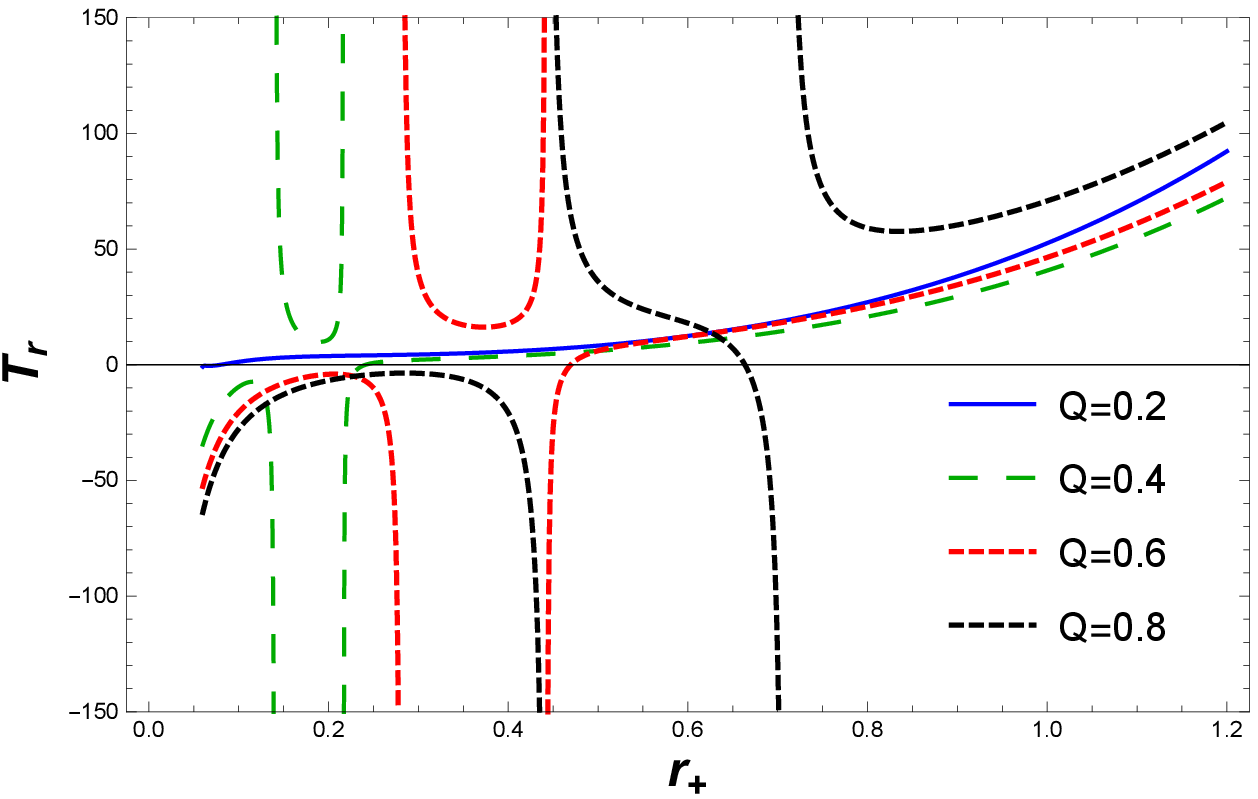}
(b)\includegraphics[width=.4\linewidth, height=2in]{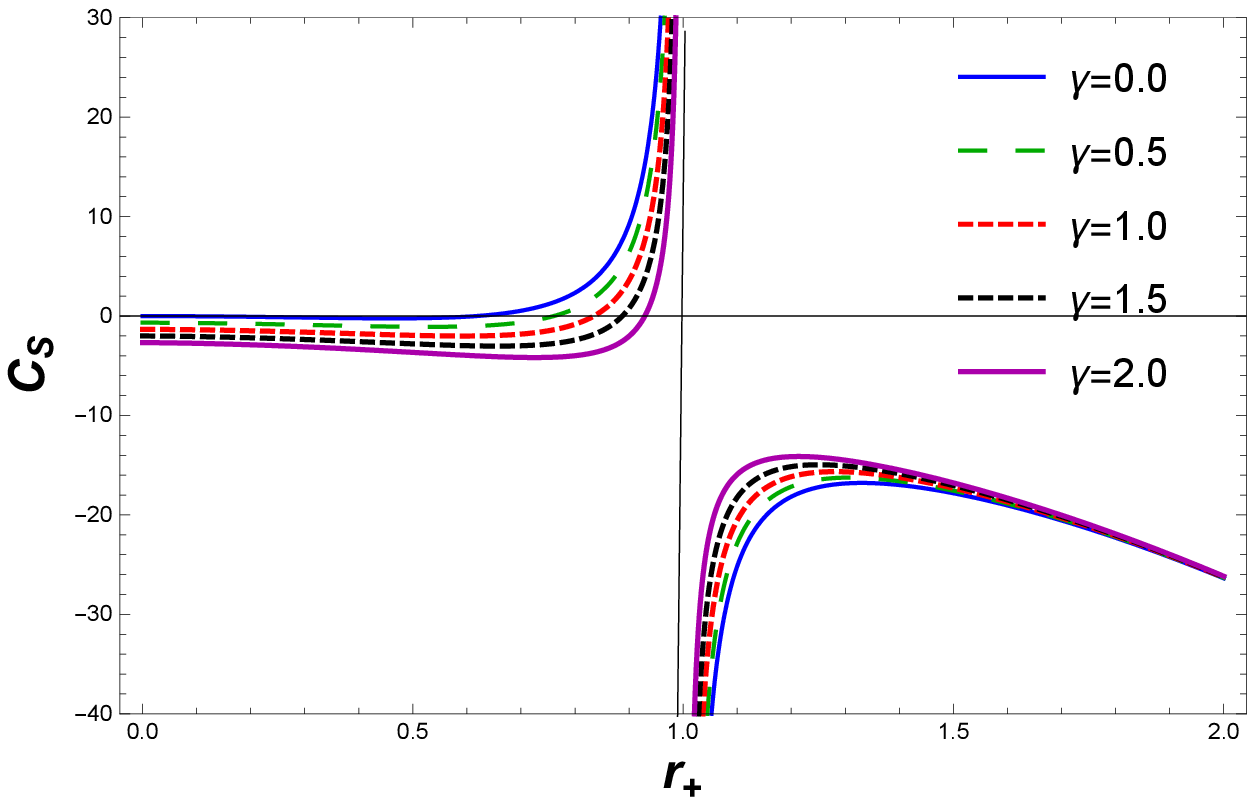}
(e)\includegraphics[width=.4\linewidth, height=2in]{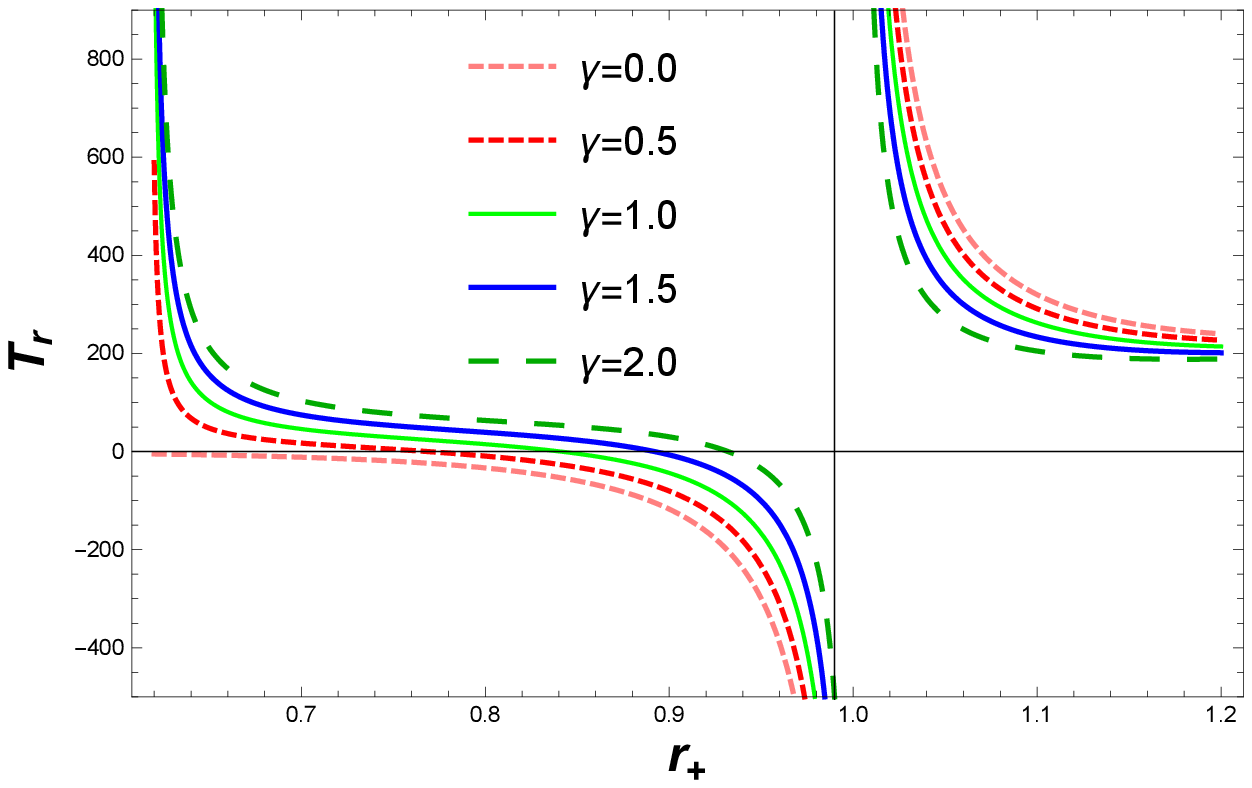}
(c)\includegraphics[width=.4\linewidth, height=2in]{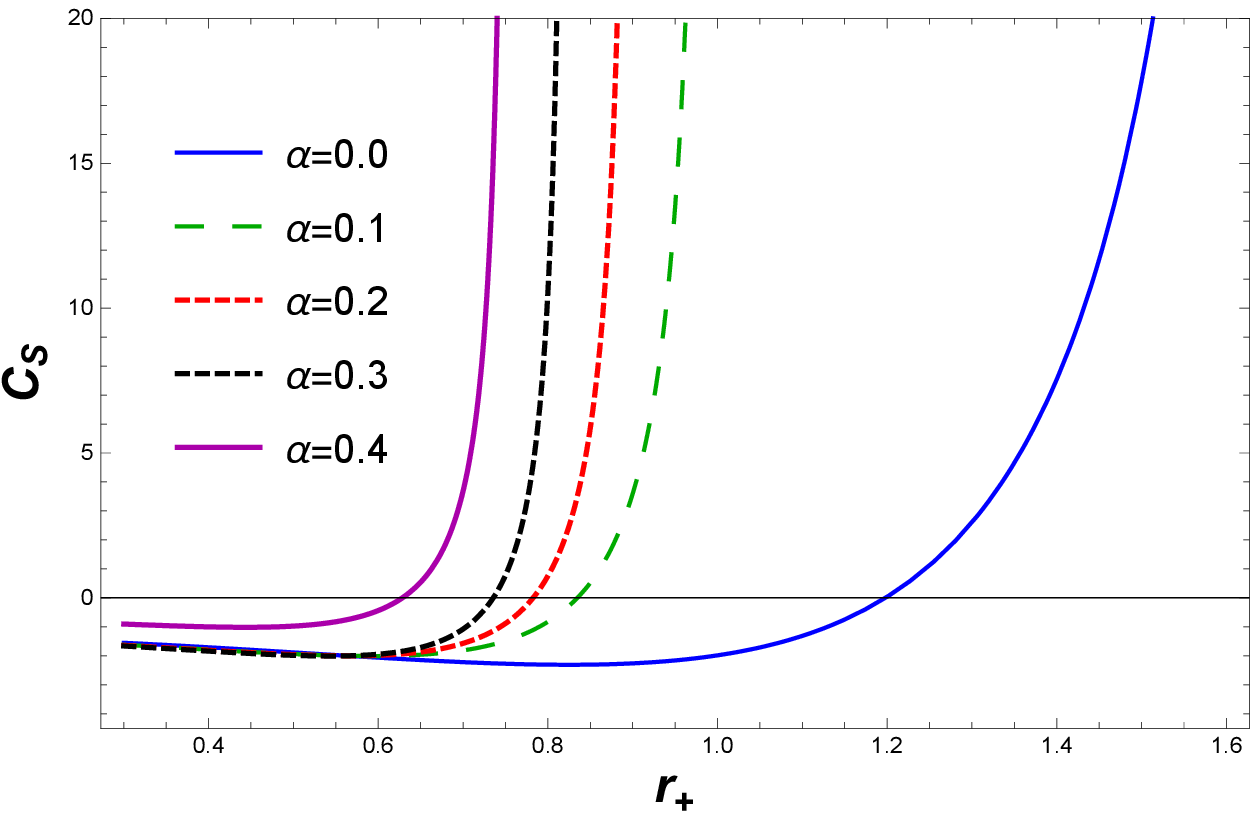}
(f)\includegraphics[width=.4\linewidth, height=2in]{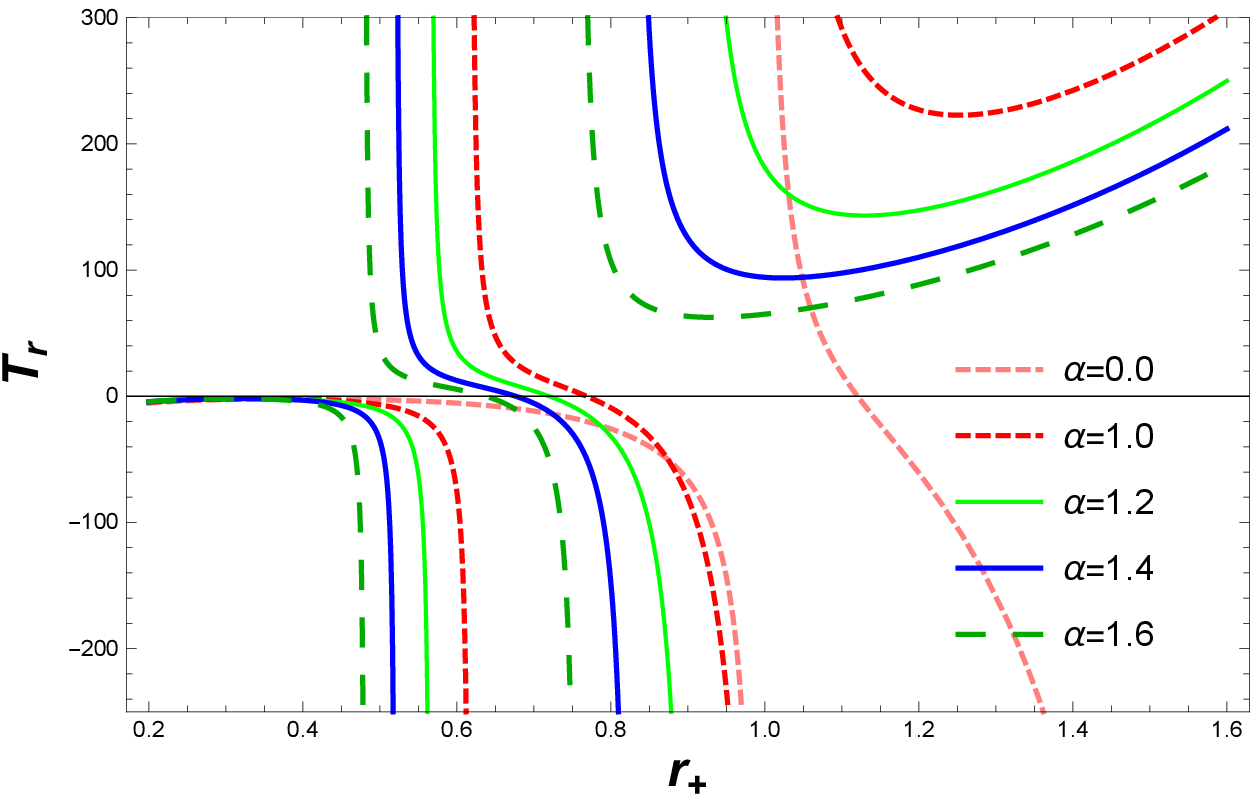}
\caption{In this Figs. (letf 3 plots) represent the graph of Corrected specific heat as a function of $r_{+}$ for $l=20$. For (a) we take  $\alpha=0.5$, $\gamma=1$, for (b) $Q=1$, $\alpha=1$  and for (c) $Q=1$, $\gamma=0.5$. Similarly (right 3 plots ) represent the graph of Trace of Hessian matrix as a function of $r_{+}$ for $l=20$. For (d) we take  $\alpha=1$, $\gamma=1$, for (e) $Q=1$, $\alpha=1$  and for (f) $Q=1$, $\gamma=0.5.$}
\end{figure}
The Fig. $8$, expresses the physical significance of specific heat and the trace of the Hessian matrix in terms of event horizon radius $r_{+}$. In Fig $8$ (left plots), we have analyzed the heat required to change the temperature of the system by varying all three parameters. There is much work that has been done \cite{14, 15, 490a, 490f, 491} relating to the heat capacity under the influence of thermal fluctuations. In Ref. \cite{14}, it is observed that phase transition of the specific heat occurs when$C_{v}\geq0$. Jawad and Shahzad \cite{15}, have proposed that the local stability of the BH system depend on the sign of specific heat, if $C_{v}\geq0$, then BH is locally stable, and becomes unstable when $C_{v}\leq0$ and phase transition happened when $C_{v}=0$.
The authors in \cite{490f}, have expressed that the heat capacity is completely unstable without the logarithmic correction and becomes stable for small BH horizon radii with the thermal fluctuations. In Fig $8$, the plot \textit{(a)}, expresses that phase transitions occur only for small values of the charge parameter and small BH horizon radius. To check the impact of the logarithmic correction due to thermal fluctuation, we focus our attention on plot \textit{(b)}, which describes the phase transition before the critical radius at $r_{c}=1$, and specific heat is divergent from the unstable to the stable region. After the critical radius, specific heat remains unstable in the given domain. From plot \textit{(c)}, it can be seen that initially heat capacity diverges from the negative region to the positive region to attain its maximum value. Finally, it can be noted that as the values of the PFDM parameter decrease, the associated specific heat becomes more stable for large BHs. To use the Hessian matrix to make sure the system is stable, we need the positive trace. For the (right plots), small BHs did not fulfil the criteria for stability while large radius BHs do in considered domain.
\section{Phase transition of temperature and heat capacity }
This study describes the Hawking temperature and heat capacity for the charged AdS BH with PFDM. By using Eqs. (\ref{a11}), and the Eqs. (\ref{a13}) give us Hawking temperature
\begin{equation}
T= \frac{l^2 \left(r_{+} (\alpha +r_{+})-Q^2\right)+3 r_{+}^4}{4 l^2 r_{+} S_0},
\end{equation}
The specific heat of the charged AdS BH with PFDM can be obtain by using Eqs. (\ref{a13}) in Eqs. (\ref{a1100}) gives
\begin{equation}
C= \frac{2 S_0 \left(l^2 \left(r_{+} (\alpha +r_{+})-Q^2\right)+3 r_{+}^4\right)}{l^2 \left(3 Q^2-r_{+} (2 \alpha +r_{+})\right)+3 r_{+}^4}.
\end{equation}
\begin{figure}[ht!]
\centering
(a)\includegraphics[width=.4\linewidth, height=2in]{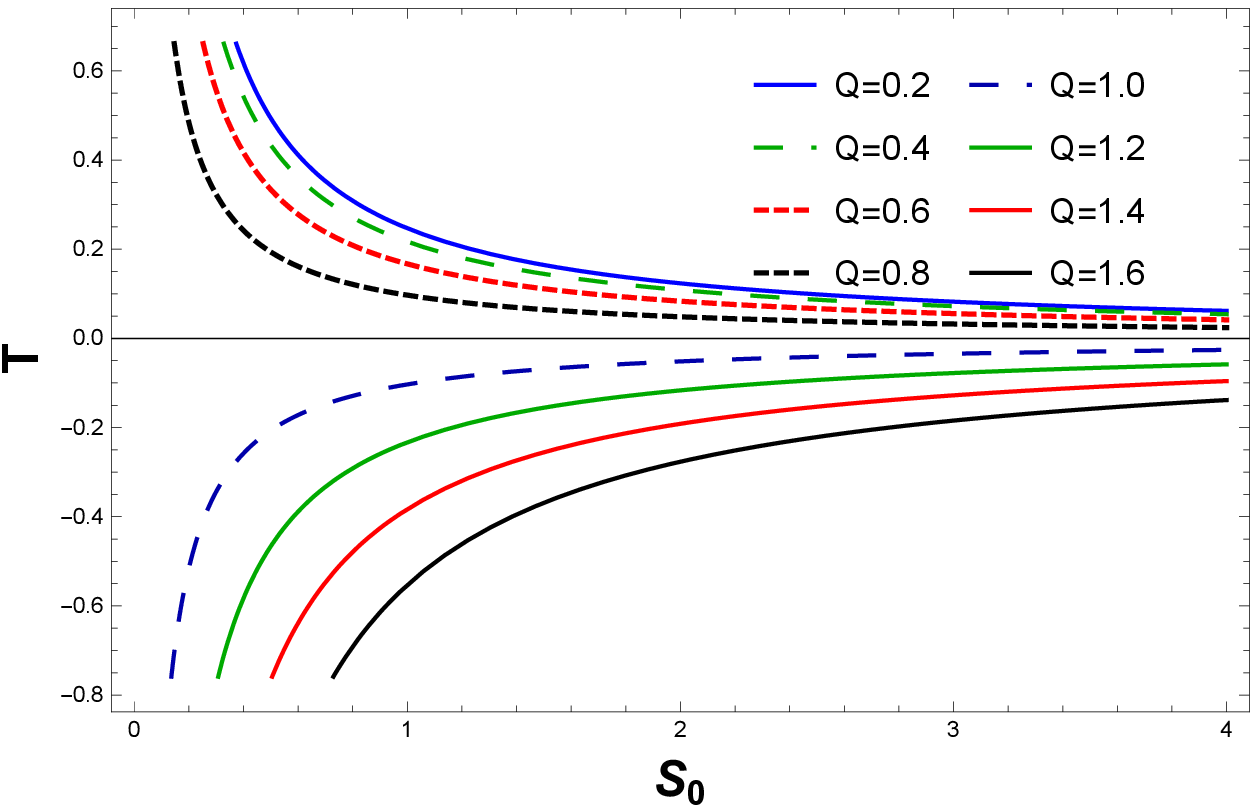}
(d)\includegraphics[width=.4\linewidth, height=2in]{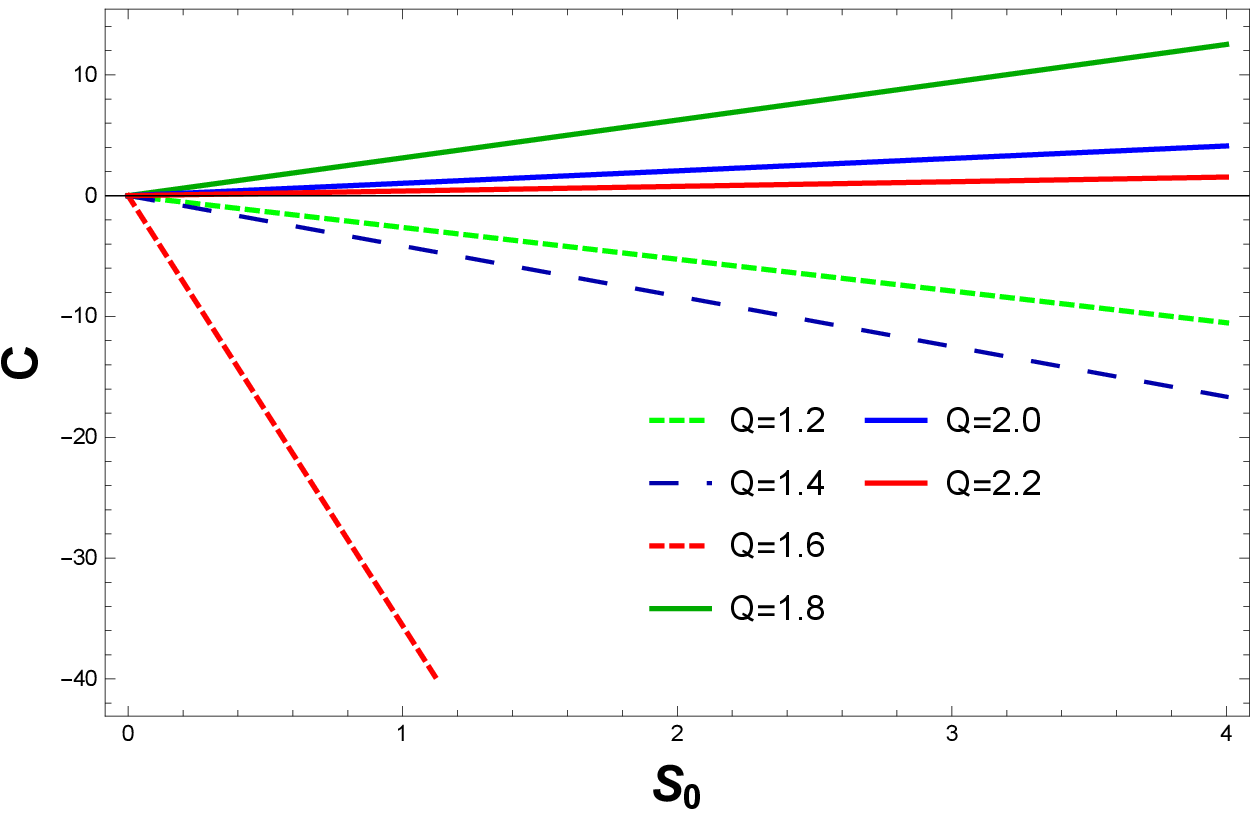}
(b)\includegraphics[width=.4\linewidth, height=2in]{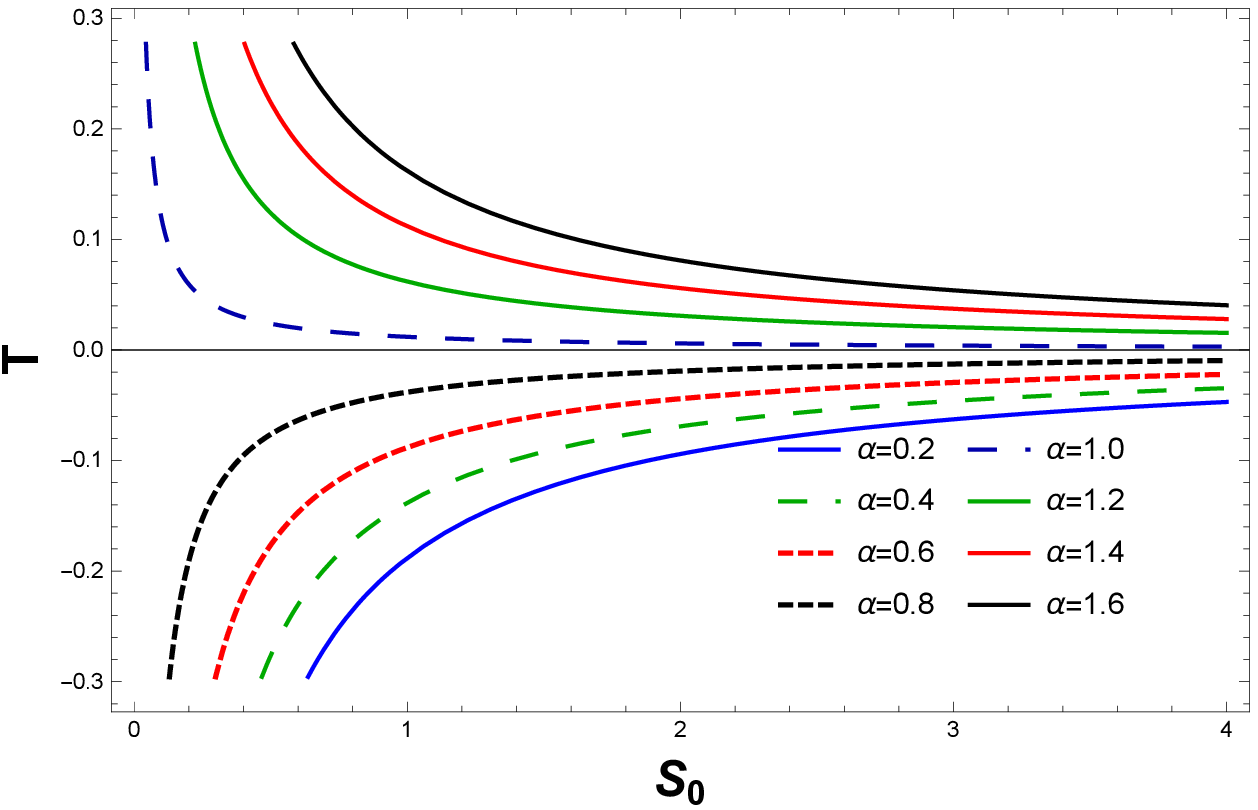}
(e)\includegraphics[width=.4\linewidth, height=2in]{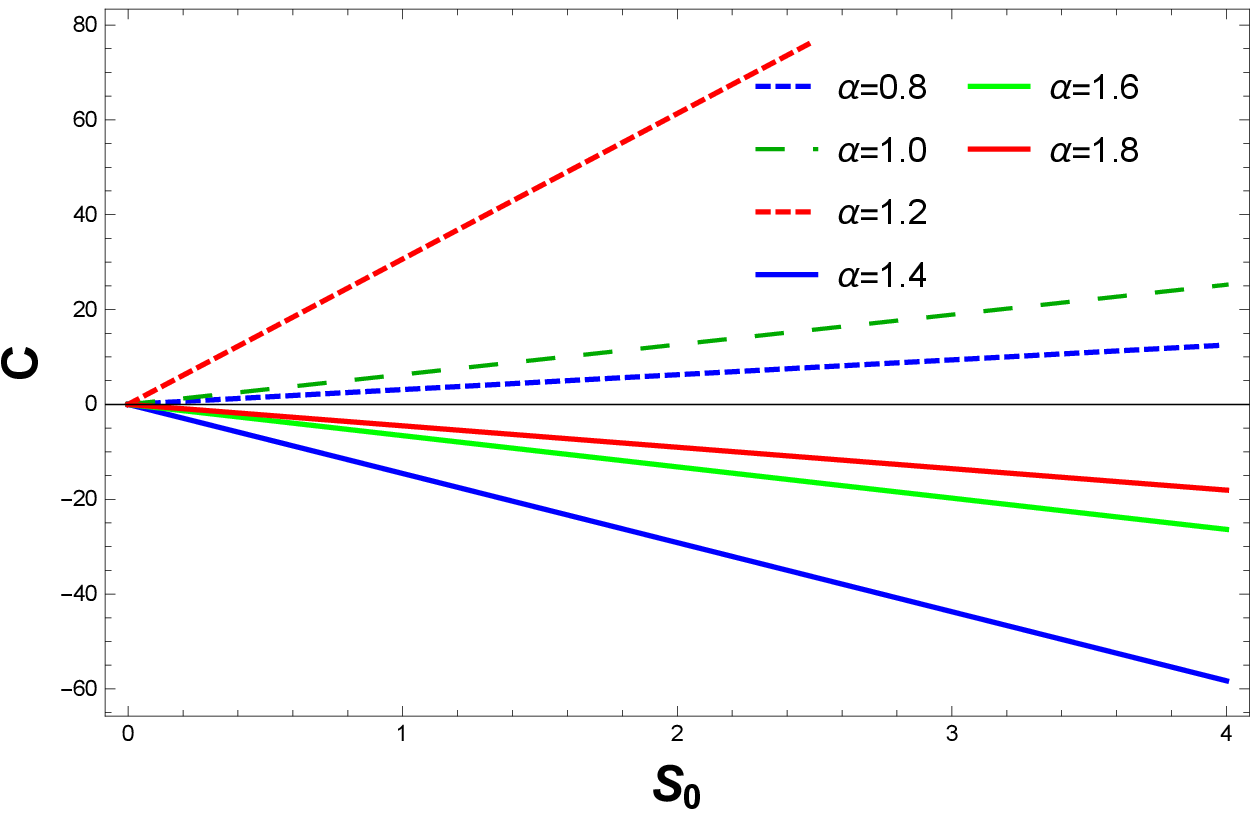}
(c)\includegraphics[width=.4\linewidth, height=2in]{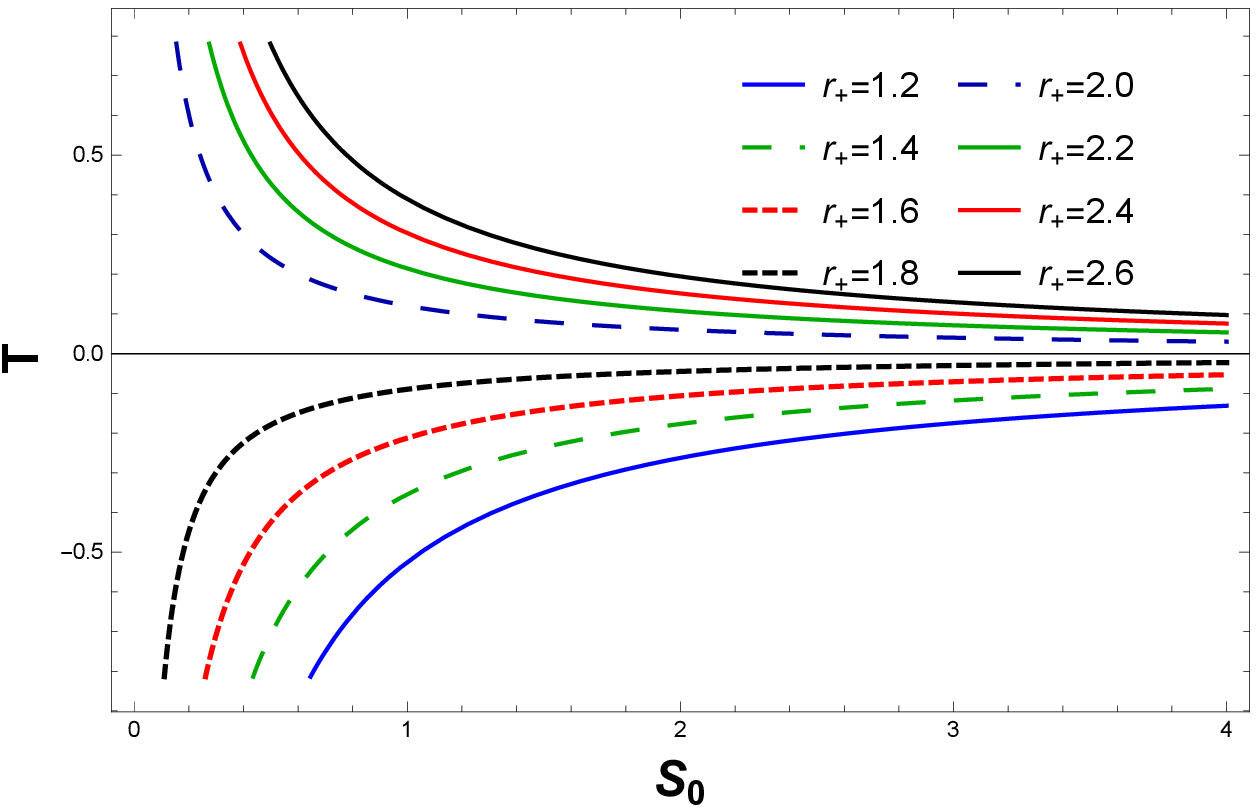}
(f)\includegraphics[width=.4\linewidth, height=2in]{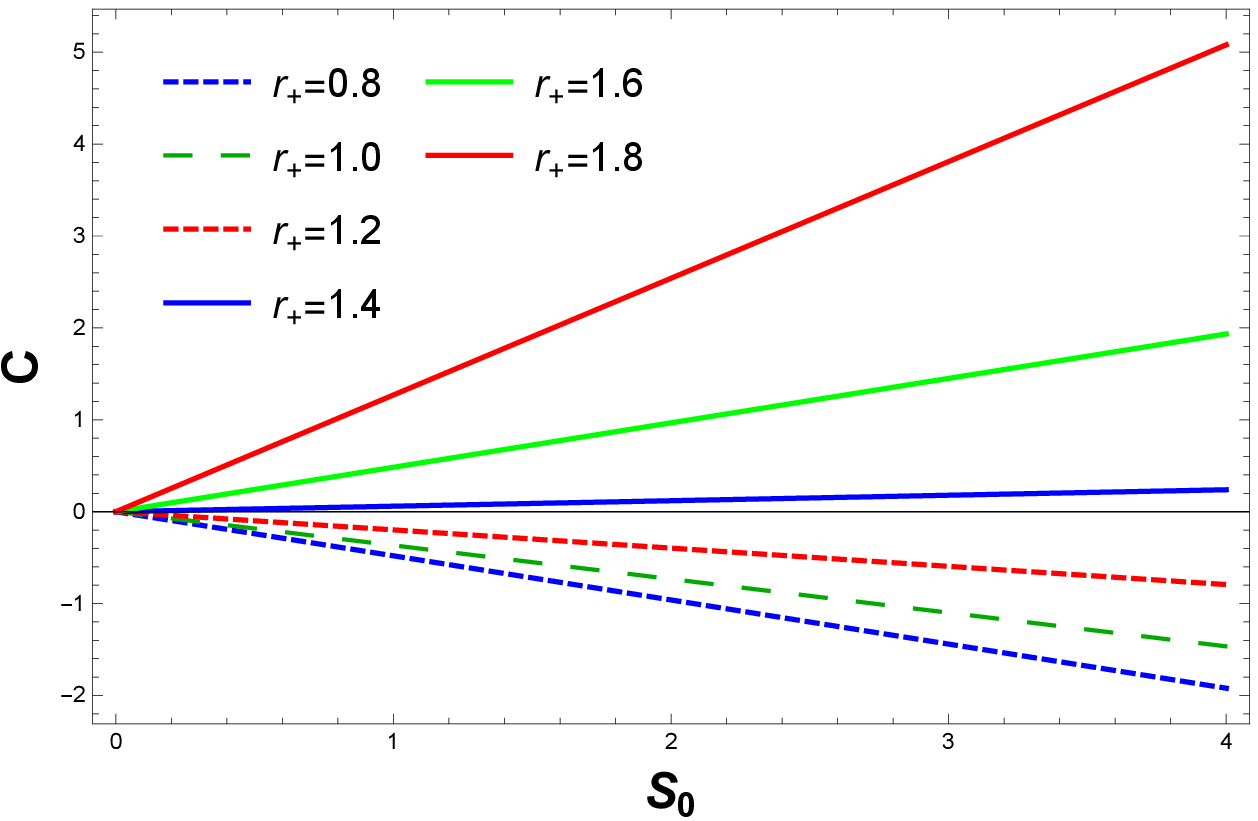}
\caption{In this fig  (letf 3 plots ) represent the graph of Temperature as a function of $r_{+}$ for $l=20$. For (a) we take  $\alpha=1$, for (b) $Q=1$. Similarly (right 3 plots ) represent the graph of specific heat as a function of $r$ for $l=20$. For (c) we take  $\alpha=1$, $Q=1$ for (d) $Q=1$ and $\gamma=0.5$.}
\end{figure}
Figure $9$ specifies the Hawking temperature (left plots) in terms of entropy. We investigate the behavior of the phase transition of the system with uncorrected entropy as done earlier in \cite{30, 51a}. In Ref. \cite{30}, authors have examined the phase change from stable region to unstable region as the values of the charge parameter increase and the Hawking temperature becomes negative for small BH as well as positive for large BH. A similar pattern appears when the PFDM parameter increases. In Hawking temperature, phase transitions occur when the charge parameter and nonlinear parameter increase as shown in \cite{51a}. Figure $9$, shows how the Hawking temperature of a RN BH surrounded by PFDM changes with entropy. It is found that Hawking temperature changes its phase from unstable (stable) to stable (unstable) for different values of the considered parameters. As the electric charge goes up in plot \emph{(a)}, the Hawking temperature of the system changes from stable to unstable. The plots \emph{(b)}, and \emph{(c)}, show opposite trends because Hawking temperature is the entropy-based phase changes in heat capacity as shown in Fig. $9$, (right plots). For different ranges of the relevant parameters, we see a phase transition in the heat capacity \cite{30, 51a}. It has been found that phase transitions occur from a stable (unstable) region to an unstable (unstable) region by increasing charge and coupling parameter values. Heat capacity remains unstable from$Q = 1.2$ to$Q = 1.6$, and its phase shifts from negative to positive for rising levels of charge at a fixed value of the PFDM parameter, as shown in plot \emph{(d)}. The phase of the plot \emph{(f)}, also changes from negative to positive as the horizon radius increases, indicating similar behavior. The various values of PFDM parameter the behavior is shown in plot \emph{(e)}.
\section{Null geodesic and quasi-normal modes}
In this section, we will look at null geodesics and the radius of the photon sphere for the charged AdS BH with PFDM. A first step towards comprehending BH thermodynamics can be attained by studying the connection between the phase transition and QNMs in the context of BH dynamics. The aim is to investigate the connection between angular velocity and the Lyapunov exponent, in a static and spherically symmetric space-time, as the angular velocity and the Lyapunov exponent of the photon sphere correspond, respectively, to the real and imaginary parts of the QNMs. We investigate the Lyapunov exponent and angular velocity with the help of radius of photon sphere. We assume the equatorial plane ($\theta=\frac{\pi}{2})$ and Lagrangian has the following form \cite{52}
\begin{equation}
\mathcal{L}=N(r)\dot{t}^2-\frac{\dot{r}^2}{N(r)}-r^2 \dot{\phi}^2,
\end{equation}
The associated components of generalized momentum ($P_{\mu}=g_{\mu\nu}\dot{x}^{\nu}-\frac{\partial\mathcal{L}}{\partial\dot{x}^{\mu}}$) are given by
\begin{equation}
P_t=E= N(r)\dot{t}= constant,\label{a44}
\end{equation}
\begin{equation}
P_r=\frac{\dot{r}}{N(r)},\label{a45}
\end{equation}
\begin{equation}
 P_{\phi }=-r^2\dot{\phi}=-L= constant,\label{a46}
\end{equation}
where we define two constants of motion, $E$ and $L$, one of which corresponds to the energy of the photon and the other to its orbital angular momentum along each geodesic.
\\
From Eqs. (\ref{a44}), and Eqs. (\ref{a46}), one can compute
\\
$ ~~~~~~~~~~~~~~~~~~~~~~~~~~~~~~~\dot{t}=\frac{E}{N(r)}$,~~ and ~~~~$ \dot{\phi}=\frac{L}{r^2}.$
\\
The associated Hamiltonian equation for the null geodesics takes the form
\begin{equation}
\mathcal{H}= N(r)\dot{t}^2-\frac{\dot{r}^2}{N(r)}-r^2 \dot{\phi}^2= E\dot{t}-L\dot{\phi}-\frac{\dot{r}^2}{N(r)}.
\end{equation}
It is simple to calculate the radial $r$-motion from the $t$-motion and the $\phi$-motion, as given below
\begin{equation}
{\dot{r}^2}+V_{eff}=0,
\end{equation}
with
\begin{equation}
V_{eff}= \frac{L^{2}}{r^2}N(r)-E^{2}.\label{a49a}
\end{equation}

There are certain unstable circular orbits of the photon spheres for the spherically symmetric static charged AdS BH with PFDM that can be found out by the following three conditions
\begin{equation}
V_{eff}=0,\label{a48}
\end{equation}
\begin{equation}
\frac{\partial V_{eff}}{\partial r}=0,\label{a49}
\end{equation}
\begin{equation}
\frac{\partial^{2}V_{eff}}{\partial r^{2}}<0.\label{a50}
\end{equation}
By using Eqs. (\ref{a49a}) in Eqs. (\ref{a49}), we get the following result
\begin{equation}
\frac{\partial V_{eff}}{\partial r}=2N(r_{ps})-r_{ps}N'(r_{ps})=0.
\end{equation}
For the respective metric function $N(r)$, we can obtain the radius of photon sphere by solving it,
\begin{equation}
4Q^{2}-6r_{ps}m+2r_{ps}^{2}-r_{ps}\alpha+3r_{ps}\alpha \log[\frac{r_{ps}}{\alpha}]=0.
\end{equation}
Finally, we need to discuss the angular velocity $\Omega$ and Lyapunov exponent $\Gamma$, of the photon sphere respectively.
Ultimately, the QNMs are associated with angular velocity $\Omega$ and Lyapunov exponent $\Gamma$, which is expressed by null geodesics as
\begin{equation}
\Omega=\frac{\sqrt{N_{ps}}}{r_{ps}},  \,\,\,\,\,\,\,\,\,\, \Gamma= \sqrt{-\frac{V''_{\text{eff}}}{2 \dot{t}^2}},
\end{equation}
which gives us
\begin{equation}
\Omega=\frac{\sqrt{{r_{ps}^{2}l^{2}}-2mr_{ps}l^{2}+Q^{2}l^{2}+r_{ps}^{4}+\alpha r_{ps} l^{2}\log[\frac{r_{ps}}{\alpha}]}}{r_{ps}^{2}l},
\end{equation}
and
\begin{equation}
\Gamma=\sqrt{\frac{\left(r_{ps} (3 \alpha +2 r_{ps})-4 Q^2\right) \left(l^2 \left(-2 m r_{ps}+Q^2+r_{ps}^2\right)+\alpha  l^2 r_{ps} \log \left(\frac{r_{ps}}{\alpha }\right)+r_{ps}^4\right)}{2 l^2 r_{ps}^6}}.
\end{equation}
\begin{figure}[ht!]
\centering
(a)\includegraphics[width=.4\linewidth, height=2in]{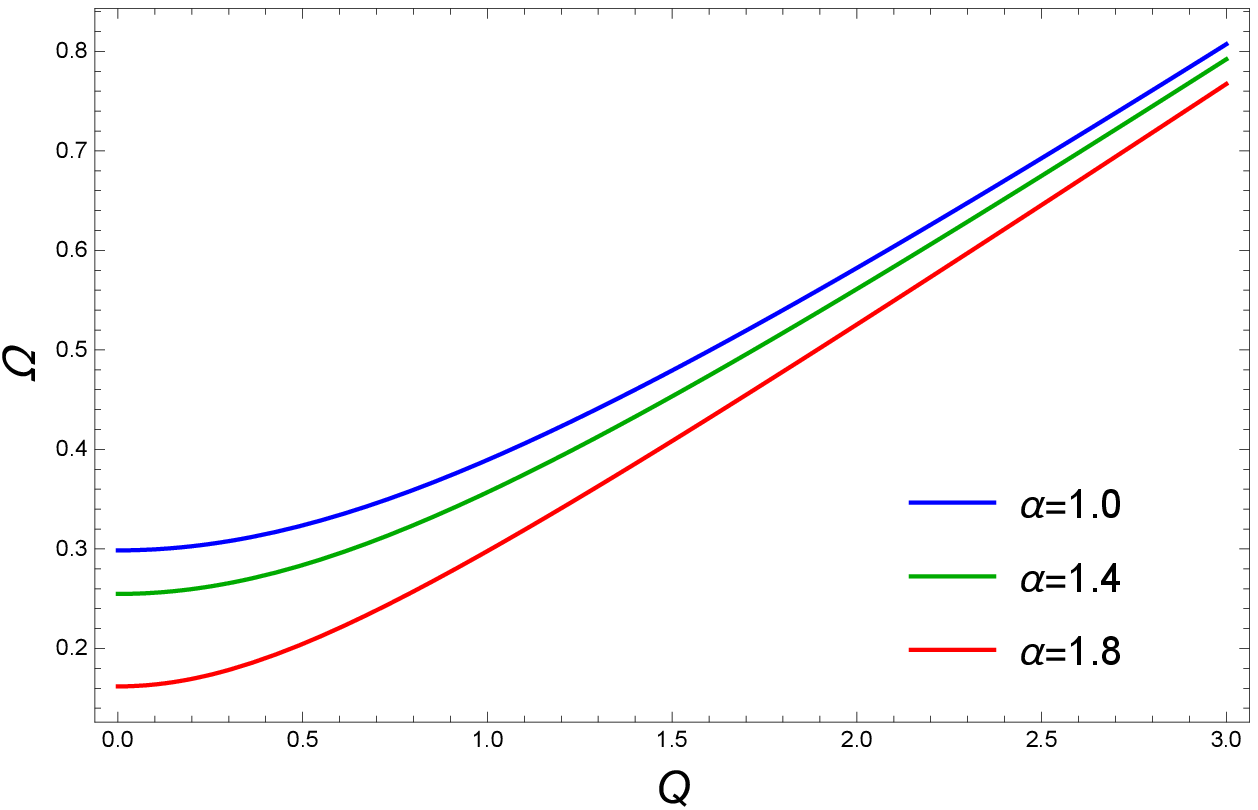}
(c)\includegraphics[width=.4\linewidth, height=2in]{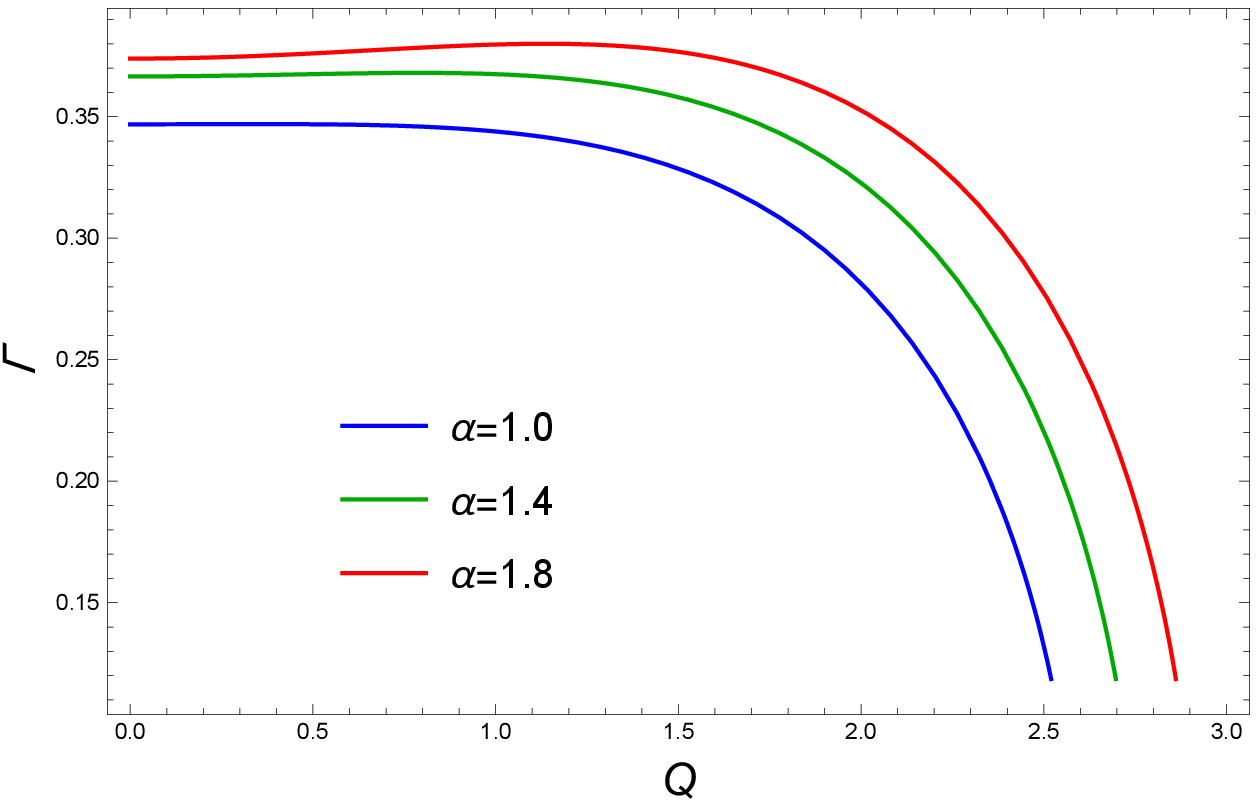}
(b)\includegraphics[width=.4\linewidth, height=2in]{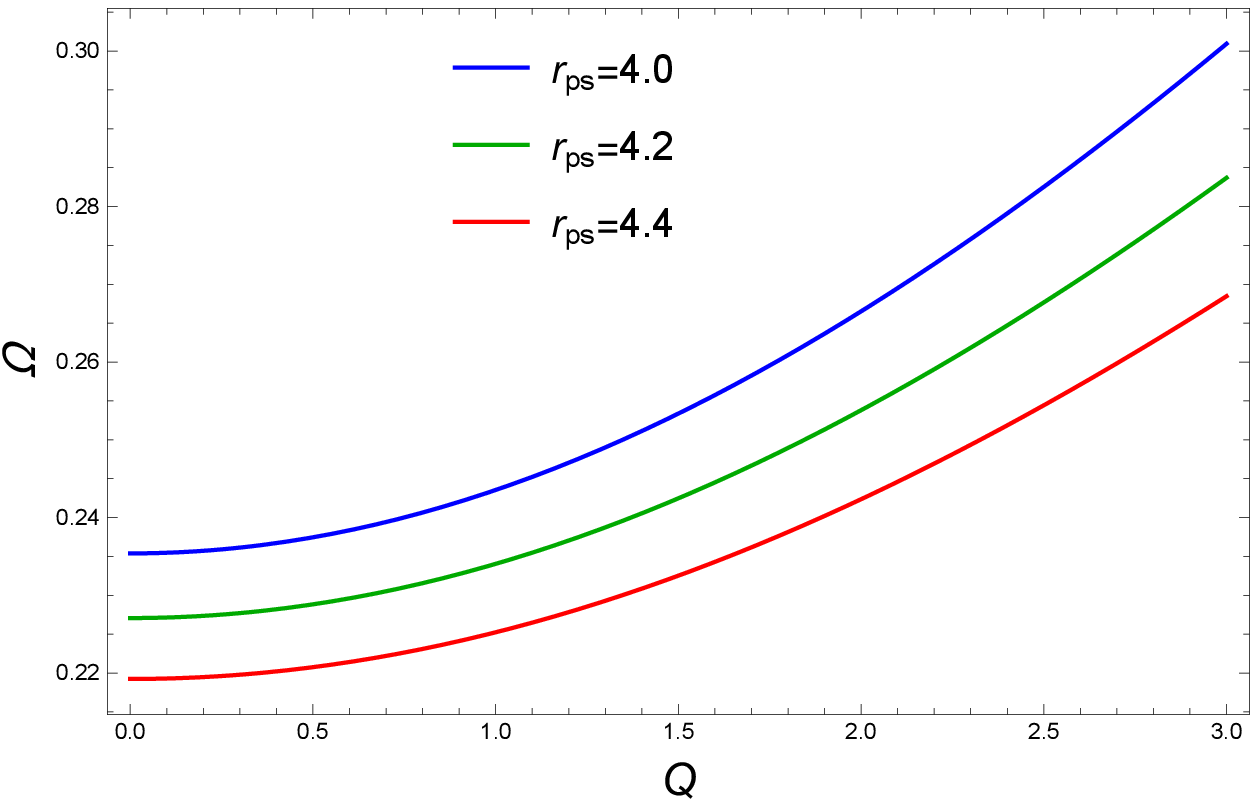}
(d)\includegraphics[width=.4\linewidth, height=2in]{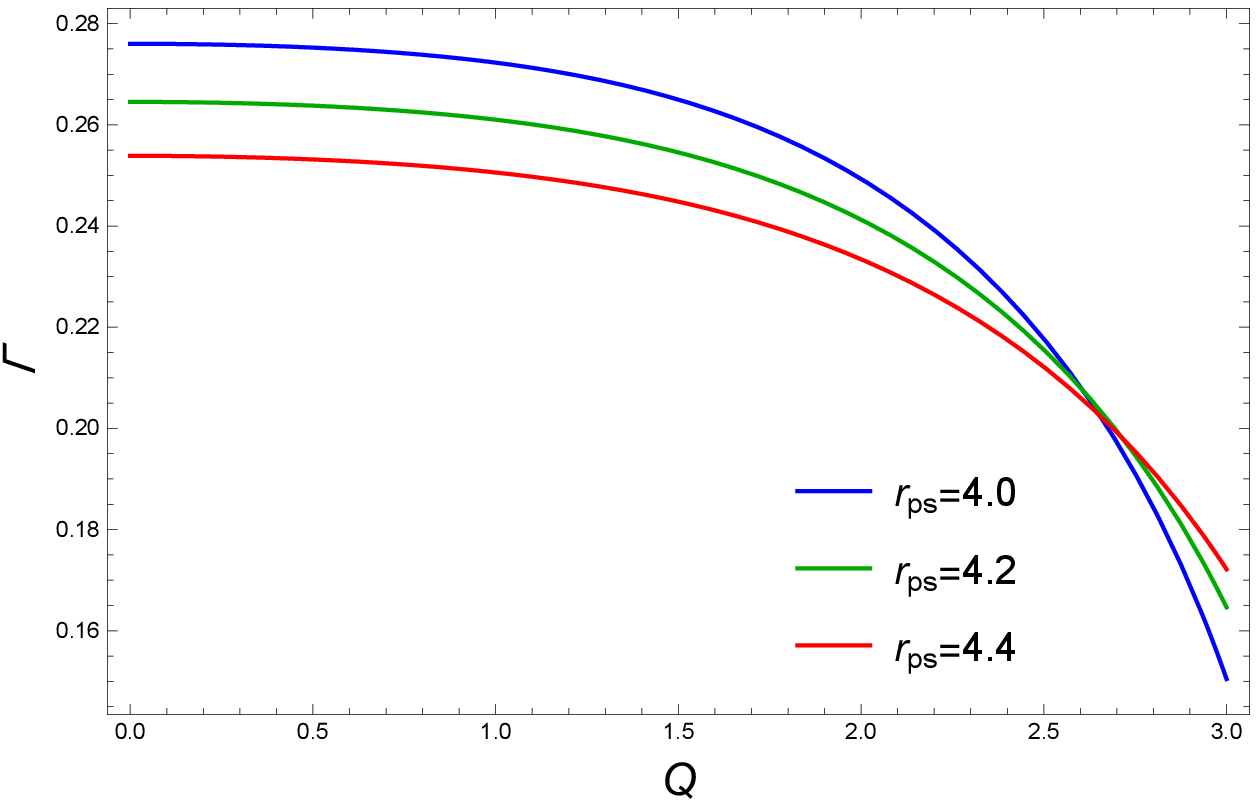}
\caption{In this Fig. (letf two plots) represent the graph of Angular velocity as a function of $Q$ with $l=20$, for plot (a), $r_{ps}=2$ and $M=1$, for plot (b), $\alpha=1$ and $M=1$. Similarly (right two plots) represent the graph of Lyapunov exponent as a function of $Q$ with $l=20$, $r_{ps}=2$ and $M=1$, for plot (c) and for plot (d), we take  $\alpha=1$ and $M=1$.}
\end{figure}\\
The physical behavior of the angular velocity as a function of charge $Q$ is shown in Fig. $10$ (left two plots). From the plot \emph{(a)}, the angular velocity increases with the values of charge. Moreover, it is clear that as the values of the PFDM parameter $\alpha$ increase, the corresponding angular velocity of the photon decreases. The angular velocity was also investigated in \cite{29, 30}, corresponding to both the radius of the photos and the coupling parameter of the considered BHs. The authors of these papers have shown that photon particles move faster with the angular velocity with higher values of $Q$, and the photons revolve more quickly with small values of the the PFDM parameter. From the plot \emph{(b)}, it is observed that the angular velocity increase to maximum as the radius of photon decreases in the considered domain. The behavior of the Lyapunov exponent, as shown in the right plots of Fig. $10$, is demonstrated with respect to two parametric values, $ \alpha$ and $r_{ps}$. From the plot \emph{(c)}, it is found that the Lyapunov exponent becomes singular rapidly for the small values of $ \alpha$ along with the small charged domain, while it shows opposite behavior for the large values of photon radius. In the literature \cite{29, 30}, the physical profile of the Lyapunov exponent was displayed and concluded that it becomes negative and decreases for large $Q$.

\section{Conclusions}
In this paper, we have observed how logarithmic corrections affect uncorrected thermodynamical parameters for a charged AdS BH surrounded by dark matter. Graphs have been used to examine the corrected thermodynamic quantities and their physical descriptions. We have analyzed Hawking temperature phase transitions and AdS BH heat capacity surrounded by PFDM. We first used first-order logarithmic corrections for a charge AdS BH with PFDM to calculate the corrected entropy, taking advantage of the fact that there are only three parameters in this model (the PFDM parameter, the charge $Q$, and the corrected parameter $\gamma$).
Because the corrected entropy is positive in all domains, we know that the system is continuous and stable. When the PFDM parameter of the Helmholtz free energy has a large value and the corrected parameter has a small value, the system is stable and energy is added quickly. It is worthwhile to mention that the validity of the modified first law of BH thermodynamics should also be discussed, and the BH's characteristics, including temperature, potential, and corrected volume, should be measured. The graphical representation of pressure indicates the unstable region for plots \emph{(a)}, and \emph{(b)}, and plot \emph{(c)}, describes the equilibrium position in the un-stable region. The enthalpy of the system expresses the stability of the system for all three parameters. Similar behavior can also be seen in Gibbs free energy in Fig. $7$ (right plots).
The system is unstable for small charge parameters, and specific heat diverges at $r_{+}=1$. Specific heat is negative before phase transition and stable thereafter, demonstrating that small-radius BHs are thermally unstable. The uncorrected Hawking temperature and uncorrected heat capacity against uncorrected entropy can also be analyzed in Fig. $9$ and their physical descriptions demonstrate the change in its phases from a positive (negative) to a negative (positive) region.
Null geodesics and QNMs were studied by finding the real component as the angular velocity and the imaginary component as the Lyapunov exponent of QNMs. The Lyapunov exponent increases rapidly with the small values of the coupling parameter, but the angular velocity of the radius of the photon decreases with increasing values of the coupling parameter $\alpha$.\\
\subsection{Appendix}
\begin{multline}
P=\Big(3 (3 l^2 Q^2-2 \alpha  l^2 r_+-l^2 r_+^2+3 r_+^4) (-32 \gamma ^2 Q^2-16 \pi  \gamma  Q^2 r_+^2+8 \pi ^2 Q^2 r_+^4+18 \alpha  \gamma ^2 r_+\\+\pi  \alpha  \gamma  r_+^3 (1-12 \log (r_+))+\pi ^2 \alpha  r_+^5(6 \log (r_+)+1)+4 \pi ^2 r_+^6){}^2 (\gamma (\log(16 \pi  l^4 r_+^4)\\-2 \log (l^2 (r_+ (\alpha +r_+)-Q^2)+3 r_+^4))+\pi  r_+^2)\Big) \Big(32 \pi  r_+^4 (\pi r_+^2-6 \gamma ) (-384 \gamma ^4 l^2 Q^4-4 \gamma ^3 l^2 r_+^2 (27 \alpha ^2 \gamma +16 \pi  Q^4)\\+3 \pi ^2 r_+^{10} (\pi  l^2 (-\pi  \alpha ^2+8 \gamma +12 \pi  Q^2)+6 \pi ^2 \alpha ^2 l^2 \log (r_+)-16 \gamma  (3 \gamma +23 \pi  Q^2))\\+384 \alpha  \gamma ^4 l^2 Q^2 r_++96 \pi  \alpha  \gamma ^3 l^2 Q^2 r_+^3 (3 \log (r_+)+8)-6 \pi ^2 \alpha  r_+^9 (12 \gamma ^2-\pi  l^2 (17 \gamma \\+6 \pi  Q^2)+2 \log (r_+)(\pi  l^2(6 \gamma +\pi  Q^2)-108 \gamma ^2))-12 \pi  \alpha  \gamma  r_+^7(192 \gamma ^2+\pi  l^2 (11 \pi  Q^2-13 \gamma )\\+2 \log (r_+) (36 \gamma ^2+\pi  l^2 (\pi  Q^2-6 \gamma )))-24 \alpha  \gamma ^2 r_+^5(9 (\pi  l^2 (2 \gamma +\pi  Q^2)-12 \gamma ^2)+10 \pi ^2 l^2 Q^2 \log (r_+))\\+12 \pi ^3 r_+^{12} (-12 \gamma +\pi  l^2+14 \pi  Q^2)-2 \pi ^2 r_+^8 (\pi  l^2 (-27 \alpha ^2 \gamma +12 \pi  Q^4+112 \gamma  Q^2)\\+9 \pi  \alpha ^2 \gamma  l^2 \log (r_+)-672 \gamma ^2 Q^2)+\pi  \gamma  r_+^6 (\pi  l^2 (117 \alpha ^2 \gamma +16 \pi  Q^4-240 \gamma  Q^2)\\+144 \pi  \alpha ^2 \gamma  l^2 \log (r_+)+4032 \gamma ^2 Q^2)-4 \gamma ^2 r_+^4 (\pi  l^2 (75 \alpha ^2 \gamma +16 \pi  Q^4-240 \gamma  Q^2)+18 \pi  \alpha ^2 \gamma  l^2 \log (r_+)+\\1440 \gamma ^2 Q^2)+6 \pi ^3 \alpha  r_+^{11} (24 \gamma +4 (\pi  l^2-27 \gamma ) \log (r_+)+\pi  l^2)+108 \pi ^4 \alpha  r_+^{13} \log (r_+)+60 \pi ^4 r_+^{14}) \Big).^{-1}
\end{multline}\label{001}

\begin{multline}
H=\frac{1}{12}\Big(\Big(3 \alpha +6 \alpha  \log(r_+)+\frac{4 \gamma  Q^2}{\pi  r_+^3}-\frac{3 \alpha  \gamma }{\pi  r_+^2}+\frac{6 Q^2}{r_+}+6 r_+-\frac{36 \gamma  r_+}{\pi  l^2}+\frac{6 r_+^3}{l^2}\\-3 (\pi  r_+^2-6 \gamma ) (-3 l^2 Q^2+2 \alpha  l^2 r_++l^2 r_+^2-3 r_+^4) (-2 \gamma  l^2 Q^2-\pi  l^2 Q^2 r_+^2+\alpha  \gamma  l^2 r_+\\+\pi  \alpha  l^2 r_+^3+r_+^4 (\pi  l^2-6 \gamma )+3 \pi  r_+^6) (r_+ (18 \alpha  \gamma ^2+\pi  r_+(r_+ (\pi  r_+ (8 Q^2+r_+ (\alpha +6 \alpha  \log (r_+)+4 r_+))\\+\alpha  \gamma  (1-12 \log (r_+)))-16 \gamma  Q^2))-32 \gamma ^2 Q^2) (\gamma (\log (16 \pi  l^4 r_+^4)-2 \log(l^2(r_+(\alpha +r_+)-Q^2)+3 r_+^4))+\pi  r_+^2)\Big)\\\Big(\pi  l^2 r_+^3 (r_+ (384 \alpha  \gamma ^4 l^2 Q^2+r_+ (r_+ (96 \pi  \alpha  \gamma ^3 l^2 Q^2 (3 \log (r_+)+8)+r_+ (r_+ (\pi  r_+ (\gamma (\pi  l^2(117 \alpha ^2 \gamma +16 \pi  Q^4-240 \gamma  Q^2)\\+144 \pi  \alpha ^2 \gamma  l^2 \log (r_+)+4032 \gamma ^2 Q^2)+r_+ (\pi  r_+(3 r_+(r_+(\pi  l^2(-\pi  \alpha ^2+8 \gamma +12 \pi  Q^2)+2 \pi  r_+(2 r_+\\(-12 \gamma +\pi  l^2+14 \pi  Q^2+\pi  r_+ (9 \alpha  \log (r_+)+5 r_+))+\alpha (24 \gamma +4 (\pi  l^2-27 \gamma) \log(r_+)+\pi  l^2))+6 \pi ^2 \alpha ^2 l^2 \log(r_+)\\-16 \gamma (3 \gamma +23 \pi  Q^2))+2 \alpha (-12 \gamma ^2+\pi  l^2(17 \gamma +6 \pi  Q^2)-2 \log (r_+)(\pi  l^2(6 \gamma +\pi  Q^2)-108 \gamma ^2)))-2(\pi  l^2 (-27 \alpha ^2 \gamma \\+12 \pi  Q^4+112 \gamma  Q^2)+9 \pi  \alpha ^2 \gamma  l^2 \log (r_+)-672 \gamma ^2 Q^2))-12 \alpha  \gamma (192 \gamma ^2+\pi  l^2 (11 \pi  Q^2-13 \gamma)\\+2 \log (r_+) (36 \gamma ^2+\pi  l^2(\pi  Q^2-6 \gamma)))))-24 \alpha  \gamma ^2(9(\pi  l^2(2 \gamma +\pi  Q^2)-12 \gamma ^2)+10 \pi ^2 l^2 Q^2 \log(r_+)))\\-4 \gamma ^2 (\pi  l^2(75 \alpha ^2 \gamma +16 \pi  Q^4-240 \gamma  Q^2)+18 \pi  \alpha ^2 \gamma  l^2 \log (r_+)+1440 \gamma ^2 Q^2)))-4 \gamma ^3 l^2 (27 \alpha ^2 \gamma +16 \pi  Q^4)))\\-384 \gamma ^4 l^2 Q^4) \Big)^{-1}\Big).
\end{multline}

\begin{multline}
G=\frac{1}{12}\Big(\Big(3 \alpha +6 \alpha  \log(r_+)+\frac{4 \gamma  Q^2}{\pi  r_+^3}-\frac{3 \alpha  \gamma }{\pi  r_+^2}+\frac{6 Q^2}{r_+}+6 r_+-\frac{36 \gamma  r_+}{\pi  l^2}+\frac{6 r_+^3}{l^2}\\-3 (\pi  r_+^2-6 \gamma ) (-3 l^2 Q^2+2 \alpha  l^2 r_++l^2 r_+^2-3 r_+^4) (-2 \gamma  l^2 Q^2-\pi  l^2 Q^2 r_+^2+\alpha  \gamma  l^2 r_+\\+\pi  \alpha  l^2 r_+^3+r_+^4 (\pi  l^2-6 \gamma )+3 \pi  r_+^6) (r_+ (18 \alpha  \gamma ^2+\pi  r_+(r_+ (\pi  r_+ (8 Q^2+r_+ (\alpha +6 \alpha  \log (r_+)+4 r_+))\\+\alpha  \gamma  (1-12 \log (r_+)))-16 \gamma  Q^2))-32 \gamma ^2 Q^2) (\gamma (\log (16 \pi  l^4 r_+^4)-2 \log(l^2(r_+(\alpha +r_+)-Q^2)+3 r_+^4))+\pi  r_+^2)\Big)\\\Big(\pi  l^2 r_+^3 (r_+ (384 \alpha  \gamma ^4 l^2 Q^2+r_+ (r_+ (96 \pi  \alpha  \gamma ^3 l^2 Q^2 (3 \log (r_+)+8)+r_+ (r_+ (\pi  r_+ (\gamma (\pi  l^2(117 \alpha ^2 \gamma +16 \pi  Q^4-240 \gamma  Q^2)\\+144 \pi  \alpha ^2 \gamma  l^2 \log (r_+)+4032 \gamma ^2 Q^2)+r_+ (\pi  r_+(3 r_+(r_+(\pi  l^2(-\pi  \alpha ^2+8 \gamma +12 \pi  Q^2)+2 \pi  r_+(2 r_+\\(-12 \gamma +\pi  l^2+14 \pi  Q^2+\pi  r_+ (9 \alpha  \log (r_+)+5 r_+))+\alpha (24 \gamma +4 (\pi  l^2-27 \gamma) \log(r_+)+\pi  l^2))+6 \pi ^2 \alpha ^2 l^2 \log(r_+)\\-16 \gamma (3 \gamma +23 \pi  Q^2))+2 \alpha (-12 \gamma ^2+\pi  l^2(17 \gamma +6 \pi  Q^2)-2 \log (r_+)(\pi  l^2(6 \gamma +\pi  Q^2)-108 \gamma ^2)))-2(\pi  l^2 (-27 \alpha ^2 \gamma \\+12 \pi  Q^4+112 \gamma  Q^2)+9 \pi  \alpha ^2 \gamma  l^2 \log (r_+)-672 \gamma ^2 Q^2))-12 \alpha  \gamma (192 \gamma ^2+\pi  l^2 (11 \pi  Q^2-13 \gamma)\\+2 \log (r_+) (36 \gamma ^2+\pi  l^2(\pi  Q^2-6 \gamma)))))-24 \alpha  \gamma ^2(9(\pi  l^2(2 \gamma +\pi  Q^2)-12 \gamma ^2)+10 \pi ^2 l^2 Q^2 \log(r_+)))\\-4 \gamma ^2 (\pi  l^2(75 \alpha ^2 \gamma +16 \pi  Q^4-240 \gamma  Q^2)+18 \pi  \alpha ^2 \gamma  l^2 \log (r_+)+1440 \gamma ^2 Q^2)))-4 \gamma ^3 l^2 (27 \alpha ^2 \gamma +16 \pi  Q^4)))\\-384 \gamma ^4 l^2 Q^4) \Big)^{-1}\Big)-\frac{1}{4 \pi  l^2 r_+^3}\Big((-l^2 Q^2+\alpha  l^2 r_++l^2 r_+^2+3 r_+^4)(\gamma  (\log (16 \pi  l^4 r_+^4)\\-2 \log (l^2 (r_+ (\alpha +r_+)-Q^2)+3 r_+^4))+\pi  r_+^2).\Big)
\end{multline}

\begin{multline}
T_{r}(H)=8\pi l^2r_{+}^3(\frac{\gamma}{l^2(r_{+}(\alpha+r_{+})-Q^2)+3r_{+}^4}+\frac{\gamma-\pi r_{+}^2}{l^2(3Q^2-r_{+}(2\alpha+r_{+}))+3r_{+}^4})\\-\frac{1}{4\pi l^2Q^2r_{+}}(2\gamma(l^2(\alpha r_{+}-3Q^2)+3 r_{+}^4) (\log(16\pi l^4r_{+}^4)-2\log(l^2(r_{+}(\alpha+r_{+})-Q^2)+3r_{+}^4))\\+\frac{1}{l^2(Q^2-r_{+}(\alpha+r_{+}))-3r_{+}^4})(2(l^4(\pi r_{+}^3(\alpha+r_{+})(r_{+}(\alpha+r_{+})-Q^2)+\gamma(2Q^2-\alpha r_{+})(3 Q^2-r_{+}(2\alpha +r_{+})))\\-3l^2r_{+}^4 (\gamma(-8Q^2+2r_{+}^2+5\alpha r_{+})+\pi r_{+}^2(r_{+}(\alpha+r_{+})-2Q^2))+18r_{+}^8(\gamma-\pi r_{+}^2))).
\end{multline}

\section*{Conflict of Interest:} The authors declare that there is no conflict of interest regarding the publication of this article.
\section*{Data Availability Statement:} The data associated with this research paper is available with the corresponding author and can be provided on request.


\begin{thebibliography}{99}
\bibitem{1} J. M. Bardeen, B. Carter and S. W. Hawking: Commun. Math. Phys. 31, 161 (1973).
\bibitem{2} J. D. Bekenstein: Phys. Rev. D 9, 3292 (1974).
\bibitem{3} S. W. Hawking: Commun. Math. Phys. 43, 199 (1975).
\bibitem{4} S. W. Hawking: Phys. Rev. D 13, 191 (1976).
\bibitem{5} R. Easther and D.Lowe: Phys. Rev. Lett. 82(1999)4967.
\bibitem{6} S. S. More: Class. Quantum Grav. 22(2005)4129.
\bibitem{7} S. Das, et al.: Class. Quantum Grav. 19(2002)2355.
\bibitem{8} M. M. Akbar and S. Das: Class. Quantum Grav. 21(2004)1383.
\bibitem{9} A. Abhay: Lectures on Non-Perturbative Canonical Gravity (World Scientic, 1991).
\bibitem{10} S. Carlip: Class. Quantum Grav. 17(2000)4175.
\bibitem{11} G. Gour, and A. J. M. Medved: Class. Quantum Grav. 20(2003)3307.
\bibitem{12} M. Alishahiha: J. High Energy Phys. 8(2007)094.
\bibitem{13} J. Sadeghi  et al.: Eur. Phys. J. C 74(2014)2680.
\bibitem{14}  B. Pourhassan and M. Fazal: Eur. phys. Lett. 111(2015)40006.
\bibitem{15} A. Jawad and M. U. Shahzad: Eur. Phys. J. C 77(2017)349.
\bibitem{16} A. Pourdarvish et al.: Int. J. Theor. Phys. 52(2013)3560.
\bibitem{17} B. Pourhassan, et al.: Gen. Relativ. Gravit. 49(2017)144.
\bibitem{18} S. Banerjee,  et al.: J. High Energy Phys. 2011(2011)143.
\bibitem{19} A. Haldar and R. Biswas, : Gen. Relativ. Gravit. 50(2018)69; Astrophys. Space Sci. 363(2018)27.
\bibitem{20} M. Zhang: Nucl. Phys. B 935(2018)170.
\bibitem{21} B. Pourhassan and S. Upadhyay, : arXiv:1910.11698.
\bibitem{22} C.V. Vishveshwara, Nature 227 (1970) 936.
\bibitem{23} J. Jing and Q. Pan: Phys. Lett. B 660(2008)13.
\bibitem{24} X. He et al.: Phys. Lett. B 665(2008)392.
\bibitem{25} R.A. Konoplya, A. Zhidenko, Rev. Modern Phys. 83 (2011) 793.
\bibitem{26} R.A. Konoplya, Z. Stuchlik, Phys. Lett. B 771 (2017) 597.
\bibitem{27} N. Breton, et al., Internat. J. Modern Phys. D 26 (2017) 1750112.
\bibitem{28} M.S. Churilova, Eur. Phys. J. C 79 (2019) 629.
\bibitem{29} M. Sharif, Z. Akhtar, Phys. Dark Univ. 29 (2020) 100589.
\bibitem{30} M. Sharif, A. Khan, Chinese J. of Phys. V. 77, June (2022), Pages 1885-1902


\bibitem{31}S. G. Ghosh et al. Ann. Phys (2021).
 \bibitem{32} A. G. Tzikas Phys. Lett. B (2019).
 \bibitem{33} D. Kastor et al. Classical Quantum Gravity (2009).
 \bibitem{34} D. Kastor et al. Classical Quantum Gravity (2010).
 \bibitem{35} B. P. Dolan Classical Quantum Gravity(2011).
\bibitem{36} E. Komatsu, K. M. Smith, J. Dunkley et al: Astro. phys. Journal, vol. 192, no. 2, p. 18, 2011.
\bibitem{37} A. G. Riess, L.-G. Strolger, J. Tonry et al. The Astro. phys. J., vol. 607, p. 665, 2004.
 \bibitem{38} R. Tharanath and V. C. Kuriakose, Mod. Phys. Lett. A 28, 1350003 (2013).
 \bibitem{39} M. Shahjalal, Nuclear Phys. B 940, 63 (2019).
 \bibitem{40} W. Javed, and R. Babar, Advances in High Energy Physics 2019, 2759641 (2019).
 \bibitem{A41}V. V. Kiselev, Class. Quantum Grav 20, 1187 (2003).
 \bibitem{A42} S. B. Chen, B. Wang and R. Su, Phys. Rev. D 77, 124011 (2008).
\bibitem{A43}B. B. Thomas, M. Saleh, and T. C. Kofane, Gen. Rel. and Grav. 44, 2181 (2012).
\bibitem{A44} S. Fernando, Mod. Phys. Lett. A 28, 1350189 (2013).
\bibitem{A45} G.-Q. Li, Phys. Lett. B 735, 256 (2014).
\bibitem{A46} Z. Xu and J. Wang, Phys. Rev. D 95, 064015 (2017).
\bibitem{A47} A, Younas, M. Jamil and S. Hussain, Phys. Rev. D 92, 084042 (2015).
\bibitem{A48} J. de Oliveira and R. Fontana, Phys. Rev. D 98, 044005 (2018).
\bibitem{A49} J. F. Navarro, C.S. Frenk, S.D.M. White, ApJ 490, 493 (1997).
\bibitem{A50} K. Dutta et al., Phys. Dark U. 100855 (2021).
\bibitem{A51} J. A Ruiz, Eur. Phys. J. Plus 136, 1 (2021).
\bibitem{A52} C. Cosme, Catarina, J. Rosab and O. Bertolamia, Quantum Theory and Symmetries: Proceedings of the 11th International Symposium, Montreal, Canada, 417(2021).
\bibitem{A53} L. Padilla et al., The Astrophys J . 909, 2 (2021).
\bibitem{A54} G. Siddhartha, T. Matos, D. Nunez, E. Ramirez, Rev. Mex. Fis. 49, 203 (2003).
\bibitem{A55} K. Saurabh, K and K. Jusufi, Eur. Phys. J. C 81, 1 (2021).
\bibitem{A56} S. Shaymatov, B. Ahmedov and M. Jamil, Eur. Phys. J. C 81, 1 (2021).
\bibitem{A57} M. Ghosh, Mod. Phys. Lett. A 36, 2150043 (2021).
\bibitem{A58} T.-C. Ma et al., Mod. Phys. Lett. A 2150112 (2021).



\bibitem{44} V. V. Kiselev: Class. Quantum. Grav. 20, 1187 (2003).
\bibitem{45} V. V. Kiselev: High Energy Phys. 11 (2003) arXiv preprint arXiv:gr-qc/0303031 (2003).
\bibitem{46} Li, M.H., Yang, K.C.: Phys. Rev. D 86, 123015 (2012)).
\bibitem{47} Liu Li Zi-Yu Fu. Hui-Ling Li Gen. Rel. Grav. 54(2022). https://doi.org/10.1007/s10714-022-02926-3.
\bibitem{47a} L.F. Abbott, S. Deser, Nucl. Phys. B 195 (1982) 76.
\bibitem{47b} S.H. Hendi, B. Esla, S. Panahiyan, A. Sheykhi, Phys. Lett. B 767 (2017) 214-225.
\bibitem{48}   P. Pradhan, Universe 5 (2019) 57.
\bibitem{49}   B. Pourhassan, K. Kokabi, Z. Sabery, Ann. Phys 399 (2018) 181.
\bibitem{490}  S. S. More,Class. Quant. Grav. 22 (2005)4129-4140.
\bibitem{490a} S. Upadhyay,  Gen. Rel. Grav. 50(2018) 128.
\bibitem{490b} B. Pourhassan, H. Farahani and S. Upadhyay, Int. J. Mod. Phys. A 34(2019).
\bibitem{490c} N. Ialam, P. A. Ganai and S. Upadhyay: J. PTEP, 2019(2019).
\bibitem{490d} M. Everton, C. Abreu and J. A. Neto, Eur. Phys. J. C. 80(2020)776.
\bibitem{490e} B. Pourhassan and S. Upadhyay: Eur. Phys. J. Plus. 136(2021).
\bibitem{490f} X. Chen, X. Huang, J. Chen and Y. Wang: Gen. Rel. Grav. 53(2021).
\bibitem{490g} O. Okcu and E. Aydiner: Phys. gen-Ph, (2017).
\bibitem{490h} A. Jawad: Class. Quantum Grav. 37(2020)185020.
\bibitem{490i} S. Pal and R. Biswas: Int. J. Theo. Phys. 61(2022).
\bibitem{491}  S. Upadhyay, S. H. Hendi, S. Panahiyan, B.E. Panah, Prog of Theor. Exp. Phys, 2018,  (2018), 093E01.
\bibitem{490j} A. Jawad, S. Chaudhry and K. Jusufi: Iran J. Sci. Technol. Trans,46(2022).
\bibitem{50} B. Pourhassan, M. Faizal, Nuclear Phys. B 913 (2016) 834.
\bibitem{51} S. Upadhyay et al.: Phys. Rev. D 95(2017)106014.
\bibitem{51a} M. Sharif, A. Khan: Chinese J. of Phys. V. 77(2022), Pages 1130-1144.
\bibitem{52} V. Cardoso, et al.: Phys. Rev. D 79 (2009) 064016.
\bibitem{53} Z. Xu, et al.: Advanc. High Ener.  Phys. vol. 2019 (2019).[ https://doi.org/10.1155/2019/2434390]

\end{thebibliography}
\end{document}